\documentclass[referee,pdflatex,sn-mathphys-num]{sn-jnl}

\usepackage{graphicx}%
\usepackage{multirow}%
\usepackage{amsmath,amssymb,amsfonts}%
\usepackage{amsthm}%
\usepackage{mathrsfs}%
\usepackage[title]{appendix}%
\usepackage[dvipsnames]{xcolor}%
\usepackage{textcomp}%
\usepackage{manyfoot}%
\usepackage{booktabs}%
\usepackage{algorithm}%
\usepackage{algorithmicx}%
\usepackage{algpseudocode}%
\usepackage{listings}%
\usepackage{subcaption} 
\usepackage{amssymb} 

\begin{document}

\title[Article Title]{CFD simulation of a Rushton turbine stirred-tank using open-source software with critical evaluation of MRF-based rotation modeling}


\author*[1]{\fnm{Alfred} \sur{Reid}}\email{areid@red-fluid.com}

\author[1]{\fnm{Riccardo} \sur{Rossi}}

\author[2]{\fnm{Ciro} \sur{Cottini}}

\author[2,3]{\fnm{Andrea} \sur{Benassi}}

\affil*[1]{\orgname{Red Fluid Dynamics}, \orgaddress{ \city{Cagliari}, \country{Italy}}}

\affil[2]{\orgname{Chiesi Farmaceutici S.p.A.}, \orgaddress{\city{Parma}, \country{Italy}}}

\affil[3]{\orgname{International School for Advanced Studies (SISSA)}, \orgaddress{\city{Trieste}, \country{Italy}}}


\abstract{
A critical evaluation of the impact of the Multiple Reference Frame (MRF) technique on steady RANS simulations of a Rushton turbine stirred-tanks is presented. The analysis, based on the open source software OpenFOAM, is focused on the choice of the diameter and thickness of the MRF region and on their effect on the predicted velocity field and mixing times in the tank. Five diameters of the MRF region are compared for the same operating conditions of the turbine, showing limited differences in velocity profiles, which are found in general good agreement with available experimental data. Significant differences are nonetheless found in the predicted levels of turbulence intensity within the tank, with a considerable amount of artificially generated turbulence at the boundary of the MRF region for the largest diameters. 
The impact of the different predictions of the turbulent field on the modeling of the mixing process in the tank is evaluated by simulating the release of a passive scalar, using the frozen-flow field hypothesis. The results show changes in mixing times up to a factor of three when comparing MRF regions of different size. Thus, the present investigation highlights the importance of assessing the effect of the MRF zone size on numerical results as a standard practice in RANS based simulations of stirred-tanks.
}

\keywords{CFD, Stirred tank, Rushton turbine, MRF, OpenFOAM, RANS}

\maketitle

\section{Introduction}
\label{sec:introduction}
Stirred tanks are widely used in several industrial sectors such as chemicals, pharmaceuticals and food and beverage. They play a major role when it comes to mixing or dissolution processes, and they are responsible  for a significant part of the total energy consumption \citep{jaszczur_general_2020}. There is thus the need to develop design tools able to predict accurately and optimize the performance of these devices.

Several types of stirred-tanks exist but they usually consist of one or more impeller systems (shaft, disc and blades), baffles on the inner skirt of the tank walls, and case-specific measuring or controlling devices like probes and injection systems. Even tough a stirred-tank may appear as a simple apparatus, the flow generated by the impeller and its interactions with the other components is often complex and highly unsteady \citep{sommerfeld_state_2004, ranade_efficient_1997}. More specifically, there are numerous parameters that have an influence on the flow field such as the tank geometry (impeller, blades, baffles) or the operating conditions (stirring speed, type of fluid, thermodynamics effects, chemical reactions) which increase the difficulty to predict the behaviour of a given stirred-tank setup. 
Significant experimental efforts have been made in the past decades towards understanding the flow in stirred-tanks and quantifying the mixing, dissolution and reaction phenomena taking place within the tank volume. However, experimental techniques present some inherent limitations. For example, it is usually impossible to obtain data in the whole tank volume and therefore only a limited zone is analyzed. Additionally, some methods may also perturb the flow and it is not always possible to run experiments on industrial-scale reactors leading to limited data for scale-up of a specific design. Lastly, laboratory experiments are expensive and potentially complex to run, hence the amount of iterations is often limited.

Thanks to the evolution of numerical techniques, Computational Fluid Dynamics (CFD) has become widely used to perform simulations of stirred-tanks and go beyond the limitations of experimental methods, as it provides a more efficient way to obtain qualitative and quantitative data. Both Large-Eddy-Simulation (LES) and Reynolds-Averaged Navier Stokes (RANS) simulations are reported in the literature, even though the latter is still the most commonly adopted for industrial applications \citep{jaszczur_general_2020}. In fact, even though LES is usually more accurate by solving directly a wide spectrum of scales of the flow, it also requires a significantly larger computational effort. On the other hand, RANS methods have shown to be sufficiently accurate in numerous industrial applications, where time to solution requirements are more stringent. Nonetheless, a proper assessment of potential limitations of RANS approaches, which compared to LES use a broader range of empirical models, is necessary. Moreover even if several papers have been dedicated to analyzing the impact of modeling parameters on numerical results, there is still the need for addressing some of the hypothesis and models adopted within the RANS framework.

Among the several modeling challenges posed by the simulation of stirred-tanks, the impeller rotation is one of the most crucial. Numerous methods are available to model the impeller rotation \citep{brucato_numerical_1998}, but they can be separated into two main categories: steady and unsteady. RANS-based steady methods are particularly interesting in industrial applications because of the lower computational effort, allowing for more in-depth optimization processes. Although different techniques exist such as the Impeller Boundary Condition (IBC) \citep{joshi_cfd_2011-1, kresta_flow_1993}, the Inner-Outer (IO) method \citep{brucato_numerical_1998} or the Snapshot approach \citep{ranade_computational_1996}, the most versatile and most commonly employed is the Multiple Reference Frame (MRF) approach \citep{luo_prediction_1994}. In this approach, the geometry studied is separated into a stationary frame and a rotating frame and, instead of explicitly rotating the impeller, a different set of equations is solved in each zone to account for the blades motion. The MRF approach thus requires the user to select the rotating frame dimension and although some application-specific recommendations exist, the choice of the MRF region extent is often overlooked. For example, existing studies have compared different MRF sizes for axial fans \citep{franzke_evaluation_2019, peng_strategy_2019} but no general guidelines are provided regarding the ideal MRF zone size. Nevertheless, these studies stress the importance of choosing the appropriate MRF region size as it is found to have noticeable impact on the flow. 

When reviewing the existing literature, it should be noted that most stirred-tank studies using the MRF approach do not specify the rotating frame diameter and height dimensions \citep{mittal_computational_2021}. Moreover, only few studies have compared different MRF zones sizes in agitated vessels and they all highlighted significant variations in the numerical results. 

For example, \citet{patil_cfd_2018} undertook a sensitivity study comparing six MRF zones with increasing diameter (from 1.1\textit{D} to 2.2\textit{D}, \textit{D} being the impeller diameter) and thickness (from 2.1\textit{W} to 3.3\textit{W}, \textit{W} being the blade width). Although they focused their analysis on the three velocity components profiles in the proximity of the shaft, they still found significant differences in the three velocity components profiles when comparing with experimental data. 

It is interesting to note that the smallest zone used, having the diameter of the impeller, gave the worst results. Likewise, \citet{de_la_concha_effect_2019} compared six different MRF zones, from 1.11\textit{D} to 1.43\textit{D} in diameter and from 3.1\textit{W} to 6.35\textit{W} in thickness. They found that at very low Reynolds number, based on the impeller diameter and rotation speed, the rotating frame size had a significant impact on results and recommended to increase the diameter and height of the MRF zone as the Reynolds number increases. Moreover, they report how local velocity profiles and power number were more sensitive to the MRF zone size than to the mesh resolution. More recently, \citet{kuschel_validation_2021} observed a sharp increase in power number when the MRF zone was below 1.5\textit{D} in diameter and below 1.5\textit{W} in thickness. They also found the power number decreased when the MRF diameter was over 2\textit{D}. 

Even though it appears there are no established guidelines for the MRF zone size selection in the literature, there seems to be a general consensus on some practices. For example, the work of \citet{jaszczur_general_2020} suggests the rotating zone should not be located in a region where gradients are high. Similarly, \citet{zadravec_influence_2007} suggested that the rotating frame boundary should include the point where the fluid motion shifts from accelerating to decelerating, which can be difficult to achieve in practice. Additionally, other authors recommend having the rotating frame boundary positioned midway between the blade tip and the baffles \citep{jaszczur_general_2020, coroneo_cfd_2011, oshinowo_predicting_2000}. In the literature, the diameter of the MRF zones ranges from 1.05\textit{D} \citep{duan_numerical_2019} to 2\textit{D} \citep{deglon_cfd_2006, kysela_cfd_2018}, and the thickness ranges from 1.18\textit{W} \citep{glover_modelling_2013} to 6.35\textit{W} \citep{de_la_concha_effect_2019} which emphasizes the heterogeneity of the MRF dimensions used in the existing CFD studies of stirred-tanks. 

Despite the aforementioned studies, the impact of the size of the MRF zone on the turbulent kinetic energy and mixing time predictions has never been addressed before and, to the best of our knowledge, remains an open point.  In this work, we thus present a critical evaluation of the impact of the MRF region size on the velocity field prediction as well as on the predicted turbulent kinetic energy field using steady RANS models. Moreover, mixing times are also qualitatively compared in order to highlight the consequences of MRF-based variations of turbulence and velocity fields on mixing dynamics when using a frozen-flow approach. With the objective of contributing to open research and innovation, the CFD model of the Rushton turbine is developed using the open-source software OpenFOAM, for which no existing research with a focus on the MRF approach is reported in the literature. 

\section{Computational methodology}
\label{sec:methodology}

\subsection{Stirred tank configuration}
\label{sec:sitrredTankConfiguration}
\begin{figure}[ht]
    \centering
    \includegraphics[width=0.7\linewidth]{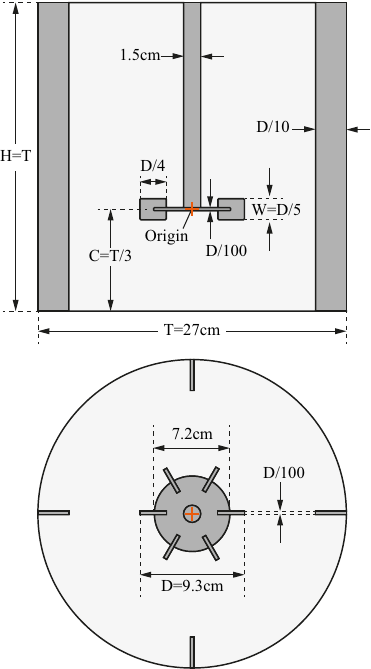}
    \caption{Dimensions of the Rushton turbine studied}
    \label{fig:geometryRushton}
\end{figure}

In this work, the geometry studied is the one used by \citet{wu_laser-doppler_1989}. It is a 27 cm baffled tank stirred with a 6-blades Rushton turbine whose dimensions are detailed in Fig. \ref{fig:geometryRushton}. 
The tank is filled with water to a height \textit{H}, the impeller diameter \textit{D} is 9.3 cm, the blade length and width are respectively $L=D/4$ and $W=D/5$ and the impeller clearance is $C=T/3$. It is important to remind that in the reference paper, the authors did not mention the shaft diameter, the thicknesses of the blades, disc and baffles, as well as the disc diameter. Therefore, these dimensions are set based on other similar studies. The disc diameter was estimated using one of the figures in the reference paper and was found to be 7.2 cm, giving a ratio to the tank height of about $T/4$. The same value was found in the literature \citep{yeoh_numerical_2004, hartmann_mixing_2006} which corroborates this measurement. For the blades, disc and baffles thickness, there is no standard value in the literature but $T/100$ has been used in several publications \citep{yeoh_numerical_2004, distelhoff_scalar_1997}. Lastly, the shaft diameter ranges from $D/4$ \citep{stoots_mean_1995} to $D/10$ \citep{malik_shear_2016} in the literature, so an intermediate value of 1.5 cm ($D/6.2$) was chosen. Fillets were added on the blades edges (0.1 mm radius) and on the disc edges (0.6 mm radius) to make the meshing process easier. The impeller and shaft are rotated in the positive direction in the coordinate system at 3.33 $s^{-1}$ (200 RPM) which corresponds to a Reynolds number of 28830 defined as: 
\begin{equation} 
\label{eq:ReynoldsNumber}
    Re=\frac{ND^2}{\nu}
\end{equation}
where \textit{N} is the rotation speed and $\nu$ the kinematic viscosity of water. Given the large extent of the experimental data available, this setup was adopted in the numerical studies of \citet{qi_numerical_2010}, \citet{malik_shear_2016}, \citet{jaszczur_experimental_2019}, and \citet{patil_numerical_2020} allowing a wide range of comparisons and validation.

\subsection{Governing equations}
\label{sec:govEquations}
The flow governing equations are solved using the ESI v2206 release \citep{opencfd_openfoam_2022} of the open-source software OpenFOAM \citep{weller_tensorial_1998}, based on the unstructured finite-volume method. The \textit{simpleFoam} solver used in the present work adopts the SIMPLE (Semi-Implicit method for Pressure Linked Equations) algorithm by \citet{caretto_two_1973} to perform steady-state calculations. The consistent variation of this algorithm (SIMPLEC) proposed by \citet{van_doormaal_enhancements_1984} is also used in some of the computations to speed-up the convergence of the numerical solution.

The impeller rotation is modeled using the MRF approach, with the impeller fixed in one position in space relative to the baffles. In the static reference frame the standard steady-state incompressible Reynolds-averaged Navier-Stokes equations are solved, more specifically the continuity equation: 
\begin{equation}
\label{eq:RANSContinuityEquation}
    \nabla \cdot \mathbf{U} = 0
\end{equation}
and the momentum equation:
\begin{equation}
\label{eq:RANSMomentumEquation}
    \nabla \cdot (\mathbf{UU}) = -\frac{1}{\rho}\nabla P+\nabla \cdot (\nu \nabla \mathbf{U}) + \nabla \cdot \mathbf{R}
\end{equation}
with $\rho$ the fluid density, $P$ the time-averaged pressure field and $\mathbf{U}$ the absolute velocity vector and $\mathbf{R}$ the Reynolds stress tensor defined as:
\begin{equation}
    \label{eq: ReynoldsStressTensor}
    \mathbf{R} = -\overline{\mathbf{U}^{'} \mathbf{U}^{'}}
\end{equation}
where $\mathbf{U}^{'}$ is the fluctuating velocity vector.

In the rotating frame, in order to account for the relative angular velocity in the MRF zone, Eq. (\ref{eq:RANSMomentumEquation}) is solved with the additional Coriolis and centrifugal terms. With a specific rearrangement of the equation that facilitates the computation, the momentum equation solved in the rotating frame is expressed as:
\begin{equation}
\label{eq:insideMRFMomentumEquation}
      \nabla \cdot (\mathbf{U_r} \mathbf{U}) = -\frac{1}{\rho} \nabla P + \nabla \cdot (\nu \nabla \mathbf{U}) -\mathbf{\Omega} \mathbf{U} + \nabla \cdot \mathbf{R}
\end{equation}
with the relative velocity vector $\mathbf{U_r}$ given by:
\begin{equation}
\label{eq:mrfVelocity}
\mathbf{U_r}=\mathbf{U}-\mathbf{\Omega} \times \textbf{r}
\end{equation}
where $\mathbf{\Omega}$ is the angular velocity vector of the rotating frame and $\textbf{r}$ is the distance from the cell centroid to the rotation axis.

The MRF method should be limited to configurations where the blades and baffle interaction is weak, a condition verified when $D < 2T$ according to \citet{oshinowo_predicting_2000}. The impact of the impeller angular position in the present setup is investigated in Appendix \ref{appendix: bladeAngularSensitivity}.

The main turbulence model used in this study is the k-$\omega$ SST proposed by \citet{menter_improved_1992, menter_two-equation_1994}
because of its versatility and generally higher-accuracy for wall-bounded flows. In the OpenFOAM release adopted in this work the following modified formulation of the turbulent kinetic energy and energy dissipation rate transport equations proposed by \citet{menter_ten_2003} are solved, written as in the OpenFOAM documentation \citep{opencfd_openfoam_2017}:
\begin{equation}
\label{eq: kEquationTurbulenceModel}
    \nabla \cdot (\rho k \mathbf{U}) = \nabla^2 (\rho D_{k} k)  + \rho G - \rho \beta^* \omega k
\end{equation} 
\begin{equation}
\label{eq: omegaEquationTurbulenceModel}
\begin{split}
        \nabla \cdot (\rho \omega \mathbf{U}) = \nabla^2(\rho D_{\omega}\omega)  + \frac{\rho \gamma G}{\nu} - \rho \beta \omega^2 - \rho (F_1 - 1)CD_{k \omega}
\end{split}
\end{equation}
with $D_k$ and $D_\omega$ the turbulent diffusion coefficients for $k$ and $\omega$ respectively, $CD_{k \omega}$ the cross diffusion term, $F_1$ the blending factor, $\beta$, $\beta^*$, $\gamma$ model constants and $G$ the production term defined as:
\begin{equation}
    \label{eq: generationTerm}
    G = \mathbf{R} \otimes \mathbf{S}
\end{equation}
where $\mathbf{S}$ is the strain rate tensor defined in Eq. (\ref{eq: StrainRateTensor}).
\begin{equation}
    \label{eq: StrainRateTensor}
    \mathbf{S} = \frac{1}{2} (\nabla \mathbf{U}+ \nabla \mathbf{U}^T)
\end{equation}

Lastly, the kinematic turbulent viscosity is defined with a viscosity limiter as:
\begin{equation}
    \label{eq: kinematicTurbulentViscosity}
    \nu_t = a_1 \frac{k}{\max (a_1 \omega_, b_1 F_{23} S)}
\end{equation}
where $\alpha_1$ and $b_1$ are model constants, $F_{23}$ is a blending factor and $S$ is magnitude of the strain rate tensor.

\subsection{Numerical mesh}
\label{mesh}
\begin{figure}
    \centering
    \includegraphics[width=0.6\linewidth]{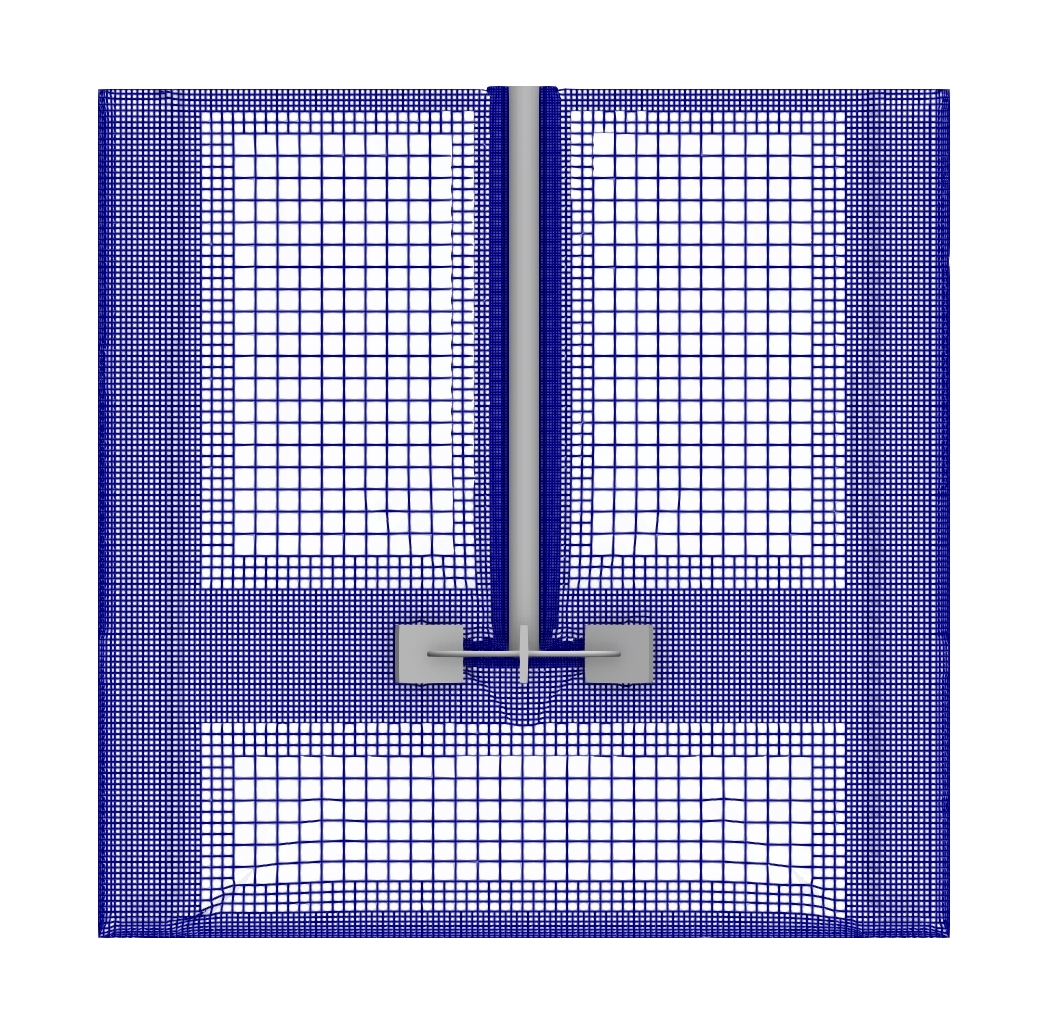}
    \includegraphics[width=0.6\linewidth]{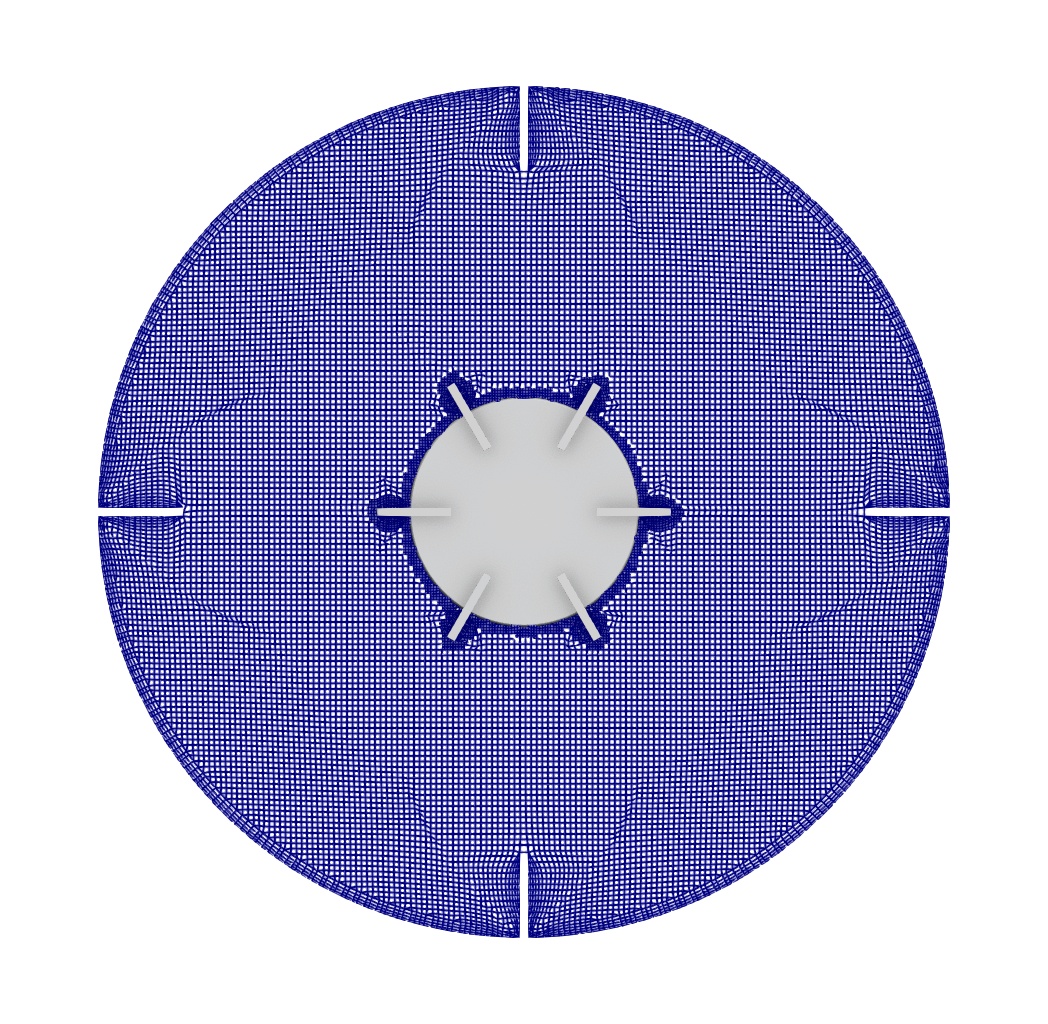}
    \caption{Example of hex-dominant grid adopted in the present study: top, side view; bottom, top view at impeller level}
    \label{fig:meshPresentation}
\end{figure}

The computational meshes adopted in the present study are created using the built-in OpenFOAM utility snappyHexMesh. Different hexahedral-dominant meshes, as the one shown in Fig. \ref{fig:meshPresentation}, are generated ranging from 1.5 to 11.8 million elements, the different mesh characteristics are detailed in Table \ref{table:meshSensitivityRecap}. A 5.76 million elements mesh was selected for the simulations after the mesh sensitivity analysis presented in Section \ref{sec:meshSensitivity} was performed. This mesh has a base cell size of 5 mm and is refined in the impeller region on a vertical distance of 1.075\textit{D} ($\sim$2 cm) above and below the impeller center line. 
The cells refined in this region are four times smaller in size than the base cells. A surface refinement was also added on all solid surfaces, with three refinement levels on the blades disc and shaft, and two levels on the tank walls, leading to cell sizes around 0.6 mm and 1.25 mm respectively. Note that the same refinement strategy has been employed with respect to the base cell size for all meshes. Lastly, in order to model the boundary layer, four prism layers are generated on all solid surfaces with an expansion ratio of 1.2. The dimensionless wall distance $y^+$ is defined as:
\begin{equation}
\label{eq:y+}
    y^+ = \frac{y u_{\tau}}{\nu}
\end{equation}
where $y$ is the wall distance and $u_{\tau}$ is the friction velocity defined as:
\begin{equation}
    \label{eq:frictionVelocity}
    u_{\tau} = \sqrt{\frac{\tau_{\omega}}{\rho}}
\end{equation}
where $\tau_{\omega}$ is the wall shear stress.
The average $y^+$ value obtained on the blades, tank and shaft surfaces for the final mesh is $y^+\sim4.13$. On the blades surface only, an average $y^+=7$ is achieved with maximum values around 35 near the blades edges and minimum values close to 0.1 near the re-circulation zones of the baffles.

\subsection{Numerical setup}
\label{sec:numerics}
For the fluid velocity field at the tank walls, baffles, disk and blades a no-slip condition has been adopted. A rotating wall velocity is applied to the shaft surface using the same rotation speed of the impeller. The free surface condition at the top of the geometry is set as symmetry as often reported in the literature \citep{qi_numerical_2010, de_la_concha_effect_2019}. With this choice it is not necessary to explicitly model the air-water interface and a single-phase solver can be adopted, which is significantly less computationally expensive than a multiphase setup. Given the range of $y^+$ values obtained in the tank, the choice of a wall modeling approach with $y^+$ insensitive wall functions is made. More specifically, the model uses the OpenFOAM boundary conditions kqRWallFunction for \textit{k}, omegaWallFunction for $\omega$, and nutUWallFunction for the turbulent viscosity $\nu_t$ on all solid surfaces.

The simulations are started using first order discretization schemes (bounded Gauss upwind) and Under Relaxation Factors (URF) of 0.4 for all quantities to ensure stability during the initial solving phase. After 250 iterations the schemes for all quantities are changed to second order (bounded Gauss linearUpwind) and at 350 iterations URF are increased to 0.7 for all quantities, except for pressure which under-relaxation is slightly reduced to a standard value of 0.3. This strategy was adopted based on a trial-and-error approach depending on the instabilities encountered and provided a satisfactory compromise between convergence speed and stability.

\begin{figure*}[ht]
  \begin{subfigure}{0.6\textwidth}
      \centering
      \includegraphics[width=0.95\linewidth]{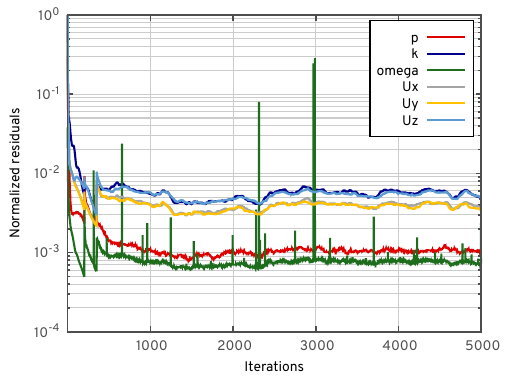}
      \caption{Residuals}
      \label{subfig:residualsKOmegaSST}
  \end{subfigure}
  \begin{subfigure}{0.6\textwidth}
      \centering
      \includegraphics[width=0.95\linewidth]{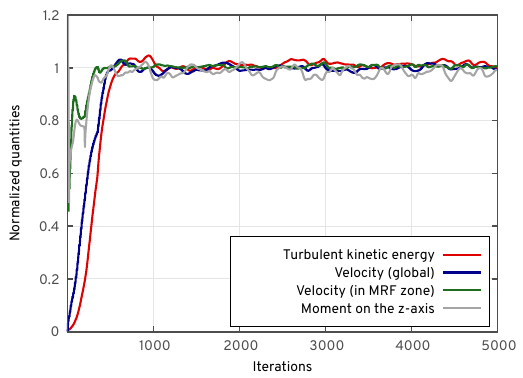}
      \caption{Volume-averaged flow quantities}
      \label{subfig:globalQuantitiesKOmegaSST}
  \end{subfigure}
  \caption{Convergence obtained with the k-$\omega$ SST turbulence model}
  \label{fig:residualsAndGlobalQuantitiesKOmegaSST}
\end{figure*}

Nevertheless, the overall convergence of the steady-state solver appears to be impacted by the underlying unsteadiness of the flow generated by the periodicity of the blade and baffles interaction as well as the fully confined flow configuration (no inlet or outlet) when switching to higher-order schemes and higher URF setup with the $k-\omega$ SST model. This is also due to the limited numerical diffusivity associated with hex-dominant meshes typically used by OpenFOAM, which are known to allow for the onset of velocity fluctuations in the flow even when solving for the steady formulation of the RANS equations.
The numerical setup was made more stable by using bounded schemes and non orthogonal correctors, nonetheless the residuals do not drop below $10^{-3}$ for velocity and for the turbulent kinetic energy $k$ and below $10^{-4}$ for pressure and the specific dissipation rate $\omega$ as shown in Fig. \ref{subfig:residualsKOmegaSST}, where the pseudo time-dependent behavior of the numerical solution can be observed.
Fig \ref{subfig:globalQuantitiesKOmegaSST} shows the evolution of the main quantities monitored with respect to the iterations: the amplitude of the oscillations after 1000 iterations is less than 5\% of the final value for all quantities and a statistically steady behavior is observed. Therefore, all the results presented in this study are averaged over the last 2000 iterations.

\subsection{Performance metrics}
\label{sec:metrics}
In this section, the main indicators selected to perform the quantitative comparison of the MRF zones and the validation of the model are presented. The power number is one of the most important parameters for a stirred tank and consists of a dimensionless number often used for scale-up purposes that quantifies the power consumption and efficiency of an impeller, defined as: 
\begin{equation}
    \label{eq:powerNumberEquation1}
    N_p = \frac{P}{\rho N^3 D^5}
\end{equation}
where $P$ is the power in Watts.
When comparing different stirring-tanks setup, low power number values indicate more efficient stirring configurations. 

There are three main methods to compute the power number in a CFD simulation. The first one consists in computing the torque by integrating the sum of the normal pressure and normal viscous forces on the impeller \citep{coroneo_cfd_2011}. The second method uses the pressure difference between the two sides of each blades to compute the torque as done by \citet{qi_numerical_2010}. The third method directly estimates the power by integrating the turbulent dissipation rate over the whole liquid volume \citep{duan_numerical_2019}.
The latter is based on the assumption that the amount of energy per unit time introduced in the reactor by the impeller is equal to the total energy dissipation rate \citep{gradov_experimentally_2017}. 
Because this specific energy balance is not rigorously verified in RANS models \citep{joshi_cfd_2011-1}, this method has shown to significantly underpredict the power number \citep{murthy_assessment_2008, lane_chapter_2000}. In this work, the first method is used and the power is calculated as follows:
\begin{equation}
    \label{eq:powerNumberEquation2}
    P=2 \pi M N
\end{equation}
where the torque around the vertical axis $M$ is computed by integrating the sum of pressure and viscous forces $F$ over each surface element $dA$ of the impeller blade surface $S$ as follows:
\begin{equation}
    \label{eq:momentEquation}
    M = \int_{S} F r \:dA
\end{equation}
where $r$ is the radial distance from the surface element to the shaft center line.

Another metric that can be used to quantify the quality of the mixing in stirred vessels is the agitation index proposed by \citet{mavros_quantification_1997}. It is a way to represent the spatially averaged mean velocity in the tank volume $\hat{U}$ as a percentage of the tip velocity $U_{tip}$ and is expressed as follows: 
\begin{equation}
    \label{eq:agitationIndexEquation}
    I_g = \frac{\hat{U}}{U_{tip}} \times 100 
\end{equation}
where 
\begin{equation}
    \label{eq:tipVelocityEquation}
    U_{tip} =  \pi ND
\end{equation}

Similar to the agitation index, the turbulence intensity is a measure of the magnitude of velocity fluctuations with respect to the flow mean velocity. It is commonly used to quantify the intensity of turbulent fluctuations in the flow and is defined as follows: 
\begin{equation}
    \label{eq:turbIntensityEquation}
    I = \frac{u^{\prime}}{\hat{U}}
\end{equation}
where $u^{\prime}$ is the root mean square of the characteristics turbulent velocity fluctuations given by:

\begin{equation}
    \label{eq:uPrimeEquation}
    u^{\prime} = \sqrt{\frac{1}{3} \left( {u^\prime_x}^2 + {u^\prime_y}^2 + {u^\prime_z}^2 \right)} = \sqrt{\frac{2}{3}k}
\end{equation}
where ${u^\prime_x}^2$, ${u^\prime_y}^2$ and ${u^\prime_z}^2$ are the components of the root mean square of velocity fluctuations.

\section{Results and analysis}
\label{sec:results}
\subsection{Mesh sensitivity study}
\label{sec:meshSensitivity}
Three different meshes with significantly different resolutions are compared to assess the impact on the numerical solution. The refinement zones dimensions and refinement levels are kept constant for all meshes and only the base cell size is changed. The base MRF zone used for the mesh sensitivity study and for the turbulence model comparison is the Zone 1 defined in Sec. \ref{sec:flowSensitivity}. The maximum element size (base cell size), minimum cell size (cell size in the impeller stream refinement region) and cell count of each mesh are detailed in Table \ref{table:meshSensitivityRecap} along with the minimum, mean and maximum $y^+$ values on the solid surfaces.

\begin{table}[ht!]
\centering
\begin{tabular}{|c|c|c|c|}
 \hline
 Mesh & Coarse & Medium & Fine \\
 \hline
 Max element size (mm)  & 7 & 5 & 4 \\
 Min element size (mm)  & 0.44 & 0.31 & 0.25 \\
 Nr. of elements (Mo) & 1.51 & 3.33 & 5.76 \\
Min $y^+$ & 0.13 & 0.11 & 0.07 \\
Average $y^+$ & 6.5 & 5.1 & 4.1 \\
Max $y^+$ & 55.3 & 39.9 & 29.9 \\
 \hline
\end{tabular}
\caption{Mesh sensitivity study parameters}
\label{table:meshSensitivityRecap}
\end{table}

The power number is often used to check the validity of CFD models for stirred-tanks, but in order to investigate the mesh sensitivity in greater details, the agitation index as well as mean turbulence intensity are also computed and compared.
According to \citet{deglon_cfd_2006}, the overall structure of the flow field and the mean velocity are not strongly influenced by the mesh size whereas the power number and the turbulent kinetic energy predictions should be more impacted. Table \ref{table:meshSensitivityPerfParamResults} shows the evolution of the mixing performance parameters for the different meshes: the relative differences are below 2.6\% for the power number and agitation index and below 4.2\% for the average turbulence intensity. When comparing the medium and fine meshes, the discrepancies drop to 0.9\% for power number and turbulence intensity and 2.2\% for agitation index. Therefore, the mesh resolution has a relatively low impact on these parameters.
\begin{table}[ht!]
\centering
\begin{tabular}{|c|c|c|c|c|} 
 \hline
 Grid & Coarse & Medium & Fine & Relative variation \\ 
  & & & & medium vs fine \\
 \hline
 Np (-) & 5.46 & 5.44 & 5.49 & 0.9\% \\ 
 Ig (\%) & 13.70 & 13.61 & 13.33 & 2.2\% \\ 
 I (\%) & 6.18 & 6.29 & 6.34 & 0.8\% \\
 \hline
\end{tabular}
\caption{Power number, agitation index, volume-average turbulence intensity obtained for the three grids studied along with the relative error between medium and fine mesh}
\label{table:meshSensitivityPerfParamResults}
\end{table}

A more stringent assessment can be made by comparing velocity and turbulent kinetic energy profiles at three radial distances from the shaft axis (namely 5, 6 and 7 cm) with the experimental results of \citet{wu_laser-doppler_1989}. In the reference experiments the velocities are measured on the 45° plane; however when using a steady-state approach in a baffled tank it is not possible to completely model or average angle-resolved features since the blades are not explicitly rotating. The simulations are thus executed at a given position of the impeller relative to the baffles and profiles of velocity and turbulent kinetic energy are averaged over the circumference of the tank from 0° to 350° by 10° steps. The velocity and turbulent kinetic energy profiles are extracted at three radial distances from the shaft (5, 6 and 7 cm) on a vertical distance of 2.5W (4.65 cm).

\begin{figure}
     \centering
     \includegraphics[width=0.55\linewidth]{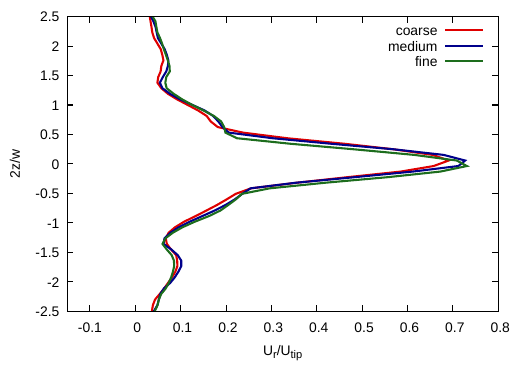}
     \includegraphics[width=0.55\linewidth]{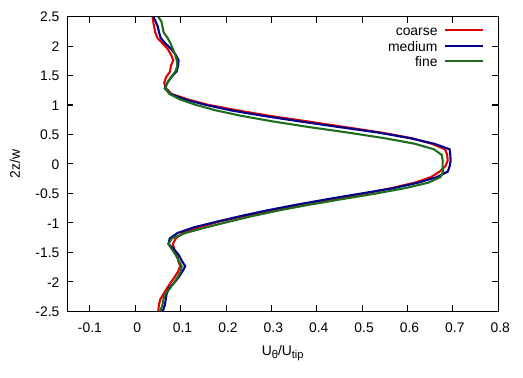}
     \includegraphics[width=0.55\linewidth]{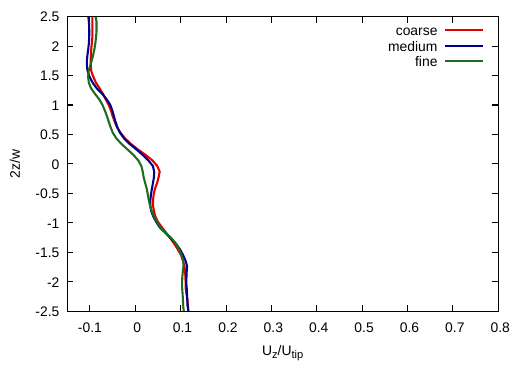}
     \caption{Effect of the grid size on predicted vertical profiles of velocity components normalized by the blade tip velocity $U_{tip}$ at 5 cm from the shaft axis: top, radial velocity $U_r$; middle, tangential velocity $U_{\theta}$; bottom, axial velocity $U_z$}
     \label{fig:velocityProfilesGridSensitivity}
\end{figure}

\begin{figure}
     \centering
     \includegraphics[width=1.0\linewidth]{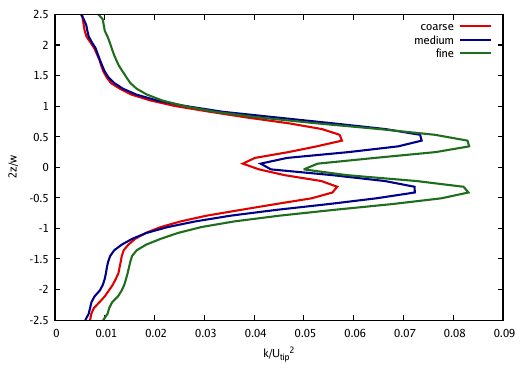}
     \caption{Effect of the mesh size on the predicted vertical profile of turbulent kinetic energy at 5 cm from the shaft axis}
     \label{fig:TKEProfilesGridSensitivity}
\end{figure}

Fig. \ref{fig:velocityProfilesGridSensitivity} shows the profiles of the three velocity components in the impeller discharge stream at a radial distance of 5 cm for the three meshes. The vertical distance $z$ is computed based on the origin of the system defined in Fig. \ref{fig:geometryRushton}, which is located at the center of the impeller.
Similarly to the observations of \citet{deglon_cfd_2006}, the velocity profiles are not significantly impacted by the mesh resolution and only small discrepancies are found between the three meshes studied (maximum deviation $\sim6$\%). Additionally, it can be observed in Fig. \ref{fig:TKEProfilesGridSensitivity} that the turbulent kinetic energy profiles, extracted at a radial distance of 5 cm, are not fully converged, despite exhibiting a converging behavior when increasing the mesh resolution. However, given the industrial-type application and the uncertainty in the reference experimental data, the finest mesh adopted in this study is deemed accurate enough for the purpose of the present work as shown in Sec. \ref{sec:results}. Similarly to the results of \citet{coroneo_cfd_2011}, the turbulence levels are higher when using a finer mesh; nonetheless the typical double peak profile is captured with all three meshes. Based on the mesh sensitivity study performed, all the results analyzed are computed using the fine mesh in the following sections of the present work.

\subsection{Turbulence model comparison}
\label{sec:turbulenceModel}
The k-$\omega$ SST model adopted in this work was developed to support wide ranges of $y^+$, a feature particularly relevant for stirred-tanks where large velocity gradients in different regions of the tank exist. However,
in the present configuration it is challenging to obtain a smooth numerical convergence and perfectly steady solution as shown in section \ref{sec:numerics}. A comparison with the k-$\epsilon$ model \citep{launder_numerical_1974, tahry_k-epsilon_1983}, which is known by CFD practitioners for promoting convergence and stability of the numerical solution, is thus performed to investigate and evaluate the partial convergence of the $k-\omega$ SST model. The model settings, boundary conditions and operating conditions are kept the same as those described in Sec. \ref{sec:numerics}.

\begin{figure*}
  \begin{subfigure}{0.6\textwidth}
      \centering
      \includegraphics[width=0.95\linewidth]{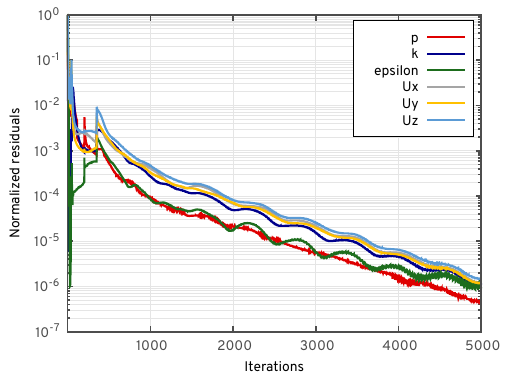}
      \caption{Residuals}
      \label{subfig:residualsKEpsilon}
  \end{subfigure}
  \begin{subfigure}{0.6\textwidth}
      \centering
      \includegraphics[width=0.95\linewidth]{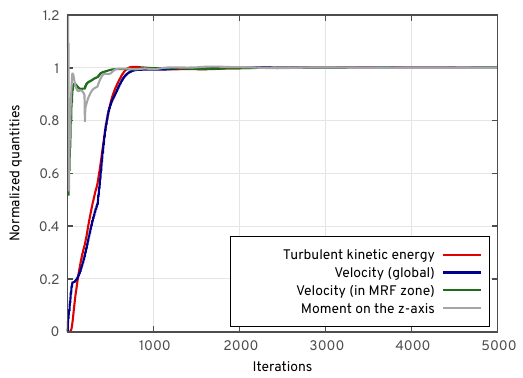}
      \caption{Volume-averaged flow quantities}
      \label{subfig:quantitiesKEpsilon}
  \end{subfigure}
  \caption{Convergence obtained with the k-$\epsilon$ turbulence model}
  \label{fig:residualsQuantitiesGlobalFigureKEpsilon}
\end{figure*}

\begin{figure}[ht]
     \centering
     \includegraphics[width=1.0\linewidth]{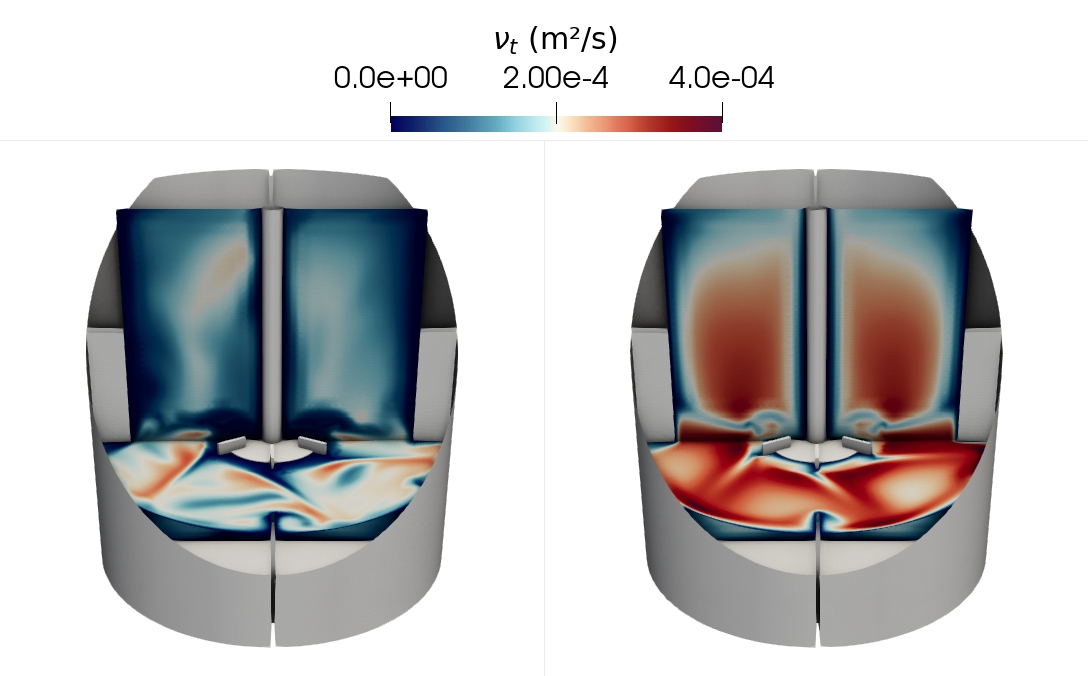}
     \caption{Values of kinematic turbulent viscosity $\nu_t$ for both turbulence models on horizontal and vertical planes: left, $k-\omega$ SST; right, $k-\epsilon$}
     \label{fig:nutCompTurbulenceModel}
\end{figure}

It is interesting to note that the instabilities discussed in Sec. \ref{sec:numerics} do not appear when using the k-$\epsilon$ turbulence model, as shown in Fig. \ref{subfig:residualsKEpsilon}. The residuals reach $10^{-6}$ after 5000 iterations, while the physical quantities in Fig. 
\ref{subfig:quantitiesKEpsilon} show no oscillation at all. The significantly higher values of kinematic turbulent viscosity obtained with the $k-\epsilon$ with respect to the $k-\omega$ SST model, shown in Fig. \ref{fig:nutCompTurbulenceModel}, illustrate the increased stability obtained with the former by dampening the potential large scale velocity fluctuations. However, the k-$\epsilon$ model comes at the cost of less flexibility for near wall treatment and a greater approximation for generally less accurate numerical solutions according to the literature, especially for turbulent quantities \citep{jaszczur_general_2020, kysela_cfd_2018,murthy_assessment_2008}. 

\begin{figure}
     \centering
     \includegraphics[width=0.8\linewidth]{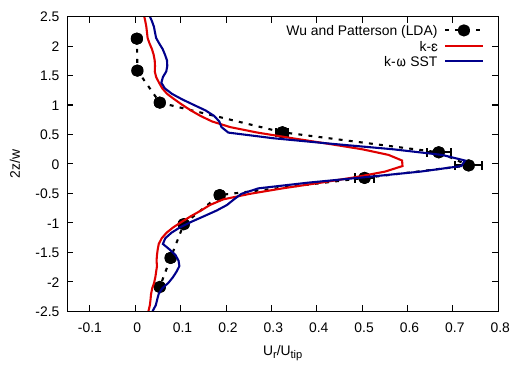}
     \includegraphics[width=0.8\linewidth]{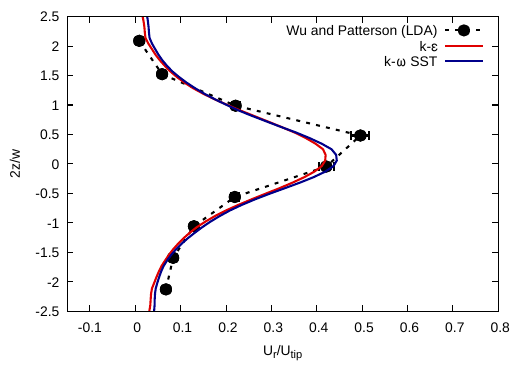}
     \caption{Computed vertical profiles of radial velocity at a distance of 5 cm (top) and 7 cm (bottom) from the shaft using the k-$\epsilon$ and k-$\omega$ SST turbulence models}
     \label{fig:UrProfilesCompTurbulenceModel}
\end{figure}

The radial velocity profiles obtained using k-$\epsilon$ and k-$\omega$ SST turbulence models at 5 cm and 7 cm away from the shaft center line are shown in Fig. \ref{fig:UrProfilesCompTurbulenceModel}.
The error bars shown for the experimental measurements of \citet{wu_laser-doppler_1989} are based on their global error estimation computed using the inflow and outflow imbalances. They estimate the uncertainty to be within 4\% for velocity and 15\% for the turbulent kinetic energy. Overall, the two models give a reasonable agreement with the experiments and very similar results for the profiles further away from the impeller. However, near the impeller the k-$\omega$ SST model better predicts the peak value of radial velocity. It is also worth noting that the under-prediction of the radial velocity when moving away from the impeller, reported by \citet{singh_assessment_2011} for both models, is not observed here. Moreover, the radial velocity is underpredicted by the k-$\epsilon$ model instead of being overpredicted as found in their study.

\begin{figure}[ht]
     \centering
     \includegraphics[width=1.0\linewidth]{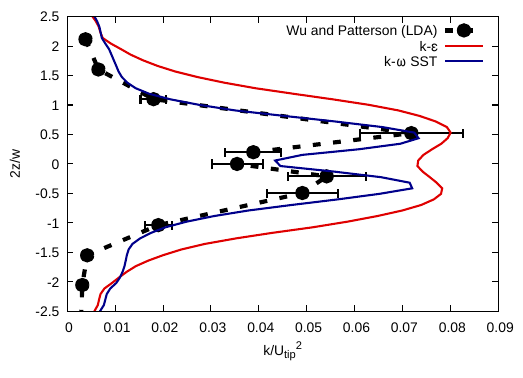}
     \caption{Computed vertical profiles of turbulent kinetic energy at a distance of 5 cm from the shaft using the k-$\epsilon$ and k-$\omega$ SST turbulence models}
     \label{fig:TKEProfilesCompTurbulenceModel}
\end{figure}

The turbulent kinetic energy profiles in the impeller discharge stream for both turbulent models are shown in Fig. \ref{fig:TKEProfilesCompTurbulenceModel} and compared to experimental data from \citet{wu_laser-doppler_1989}.
The k-$\epsilon$ model over-predicts the overall turbulent kinetic energy and does not clearly capture the typical double peak profile. On the other hand, the k-$\omega$ SST model yields a satisfactory prediction although it overpredicts the peak under the impeller center line. These results are in agreement with those of \citet{singh_assessment_2011} where it was also found that the k-$\epsilon$ model overpredicted $k$ close to the impeller and that the k-$\omega$ SST model resulted in a more accurate prediction.

\begin{figure*}[ht]
  \begin{subfigure}{1.0\textwidth}
      \centering
      \includegraphics[width=1.0\linewidth]{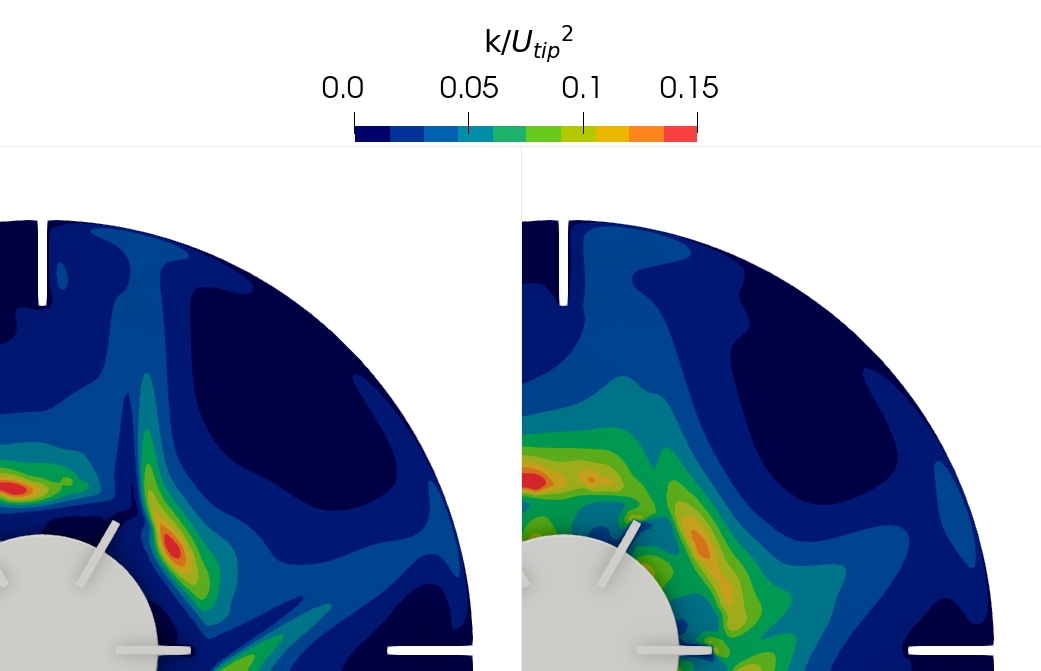}
      \caption{Horizontal plane located at 2z/w=-0.35}
      \label{subfig:TKEContoura}
  \end{subfigure}
  \begin{subfigure}{1.0\textwidth}
      \centering
      \includegraphics[width=1.0\linewidth]{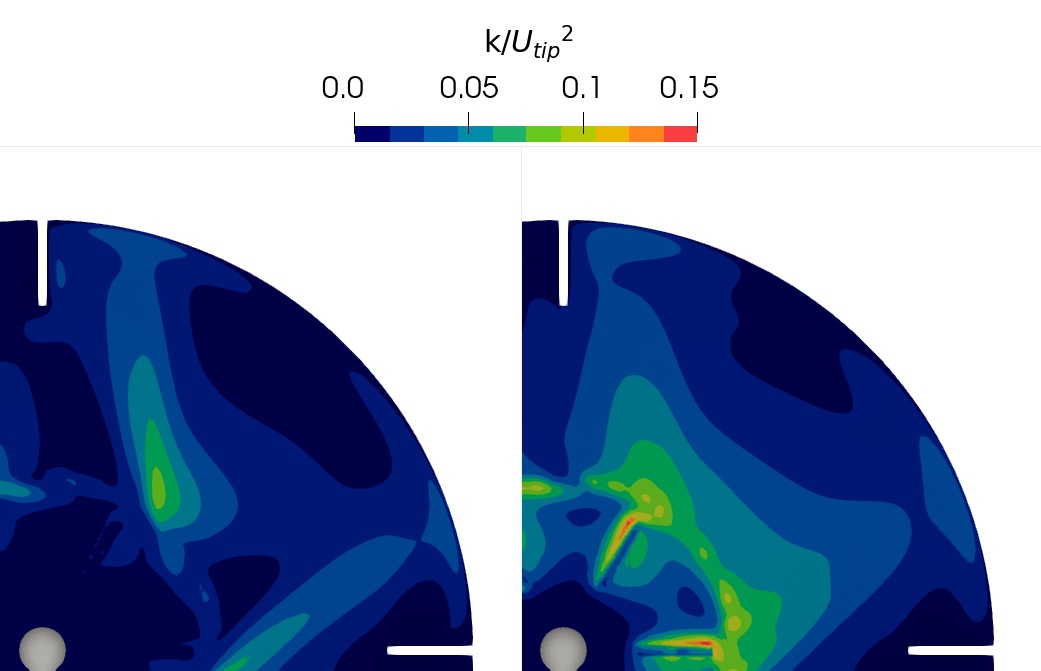}
      \caption{Horizontal plane located at 2z/w=1.0}
      \label{subfig:TKEContourb}
  \end{subfigure}
  \caption{Impact of turbulence models on the development of turbulent kinetic energy normalized by the square of the impeller tip velocity: left, k-$\omega$ SST; right, k-$\epsilon$}
  \label{fig:TKEContoursComparisonDerksen}
\end{figure*}

Fig. \ref{fig:TKEContoursComparisonDerksen} shows the turbulent kinetic energy on two different horizontal planes obtained with the $k-\omega$ SST and the $k-\epsilon$ turbulence models. The latter shows higher global levels of turbulent fluctuations and also displays a less concentrated distribution of the turbulent kinetic energy. On the other hand, while still maintaining similar basic turbulent features, the $k-\omega$ SST yields a closer agreement with the LES simulations reported in the work of \citet{derksen_large_1999}. More specifically, the turbulence levels are found significantly lower near the disc and blades on both horizontal planes, and the regions of high-turbulence generated by each blade are separated by lower turbulence zones on the plane $2z/w=-0.35$ in Fig. \ref{subfig:TKEContoura}. The better near-wall behavior of the $k-\omega$ SST model likely contributes to its improved agreement with the LES data in the vicinity of the disc and blades. Nonetheless, as one can expect, none of the two turbulence models is able to fully predict all the flow features obtained in the LES reference data as the present results are computed with a steady RANS approach.

Despite showing a less optimal convergence behavior compared to the $k-\epsilon$ model, the results obtained with the $k-\omega$ SST are in good agreement with the reference data and are consistent with the $k-\epsilon$ results. Based on these results, the $k-\omega$ SST turbulence model is considered to have sufficient robustness to effectively support the MRF zone comparison of the present work.

\subsection{MRF zone size analysis}
\label{sec:MRFZone}
\subsubsection{Impact on the flow field}
\label{sec:flowSensitivity}
\begin{figure}
     \centering
     \includegraphics[width=1.0\linewidth]{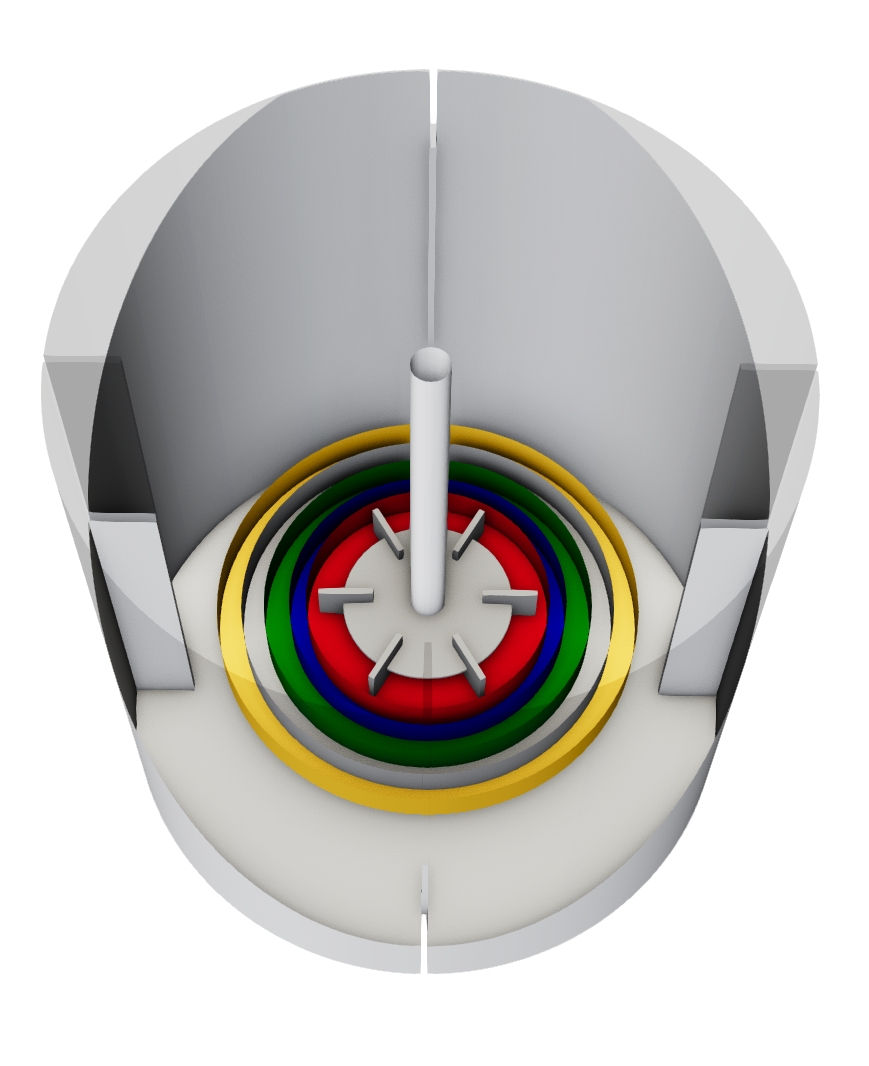}
     \caption{Size of the different MRF zones compared in this study ranging from 1 to 5, from the smallest to the largest zone}
     \label{fig:MRFDiameterOutlines}
\end{figure}

In order to assess the influence of the MRF zone on the predicted flow field, five different zones are compared with diameters of 1.10\textit{D}, 1.26\textit{D}, 1.49\textit{D}, 1.70\textit{D} and 1.93\textit{D} with a constant thickness of 1.55\textit{W}. In order to facilitate their identification, an ID is assigned to each zone in Fig. \ref{fig:MRFDiameterOutlines}, ranging from 1 to 5, from the smallest to the largest zone. The volume of the smallest zone adopted in this work is slightly greater than the volume swept by the impeller, thus limiting the rotation effect to the impeller vicinity. The increasing larger zones extend farther in the tank volume up to Zone 5, having the largest diameter and being the closest to the baffles location. Most studies reported in the literature use MRF regions with dimensions within those of Zone 3 and Zone 5 \citep{lane_chapter_2000, coroneo_cfd_2011, kysela_cfd_2018, jaszczur_general_2020, mittal_computational_2021}.

\begin{figure}[ht]
     \centering
     \includegraphics[width=1.0\linewidth]{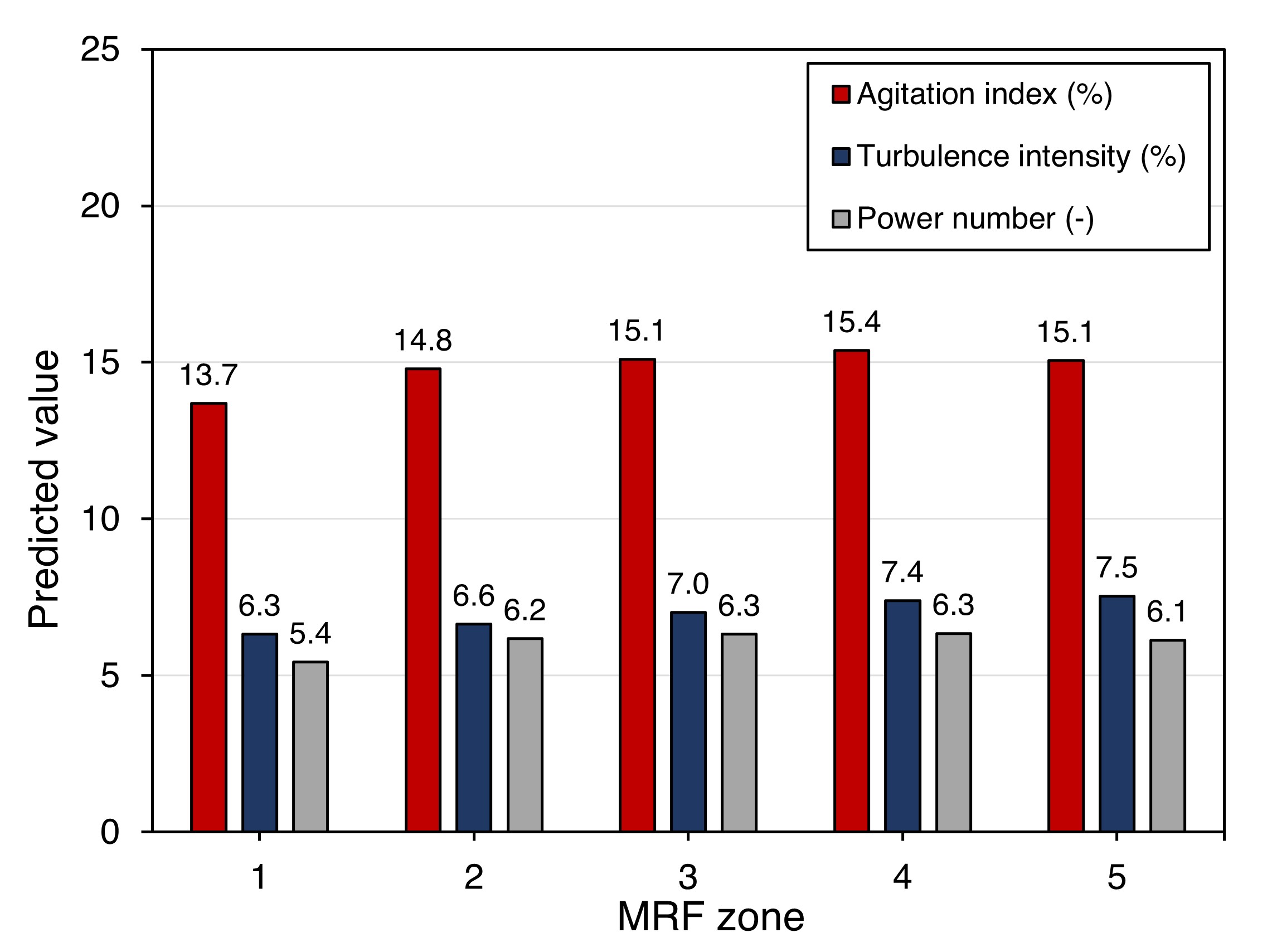}
     \caption{Effect of the MRF zone on the power number, agitation index and average turbulence intensity in the tank}
     \label{fig:MRFDiameterSensitivityPerfParam}
\end{figure}

\begin{figure}
     \centering
     \includegraphics[width=0.55\linewidth]{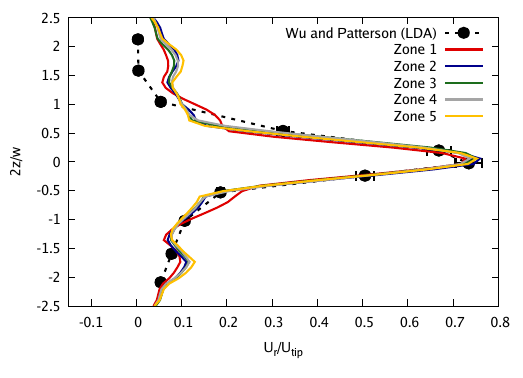}
     \includegraphics[width=0.55\linewidth]{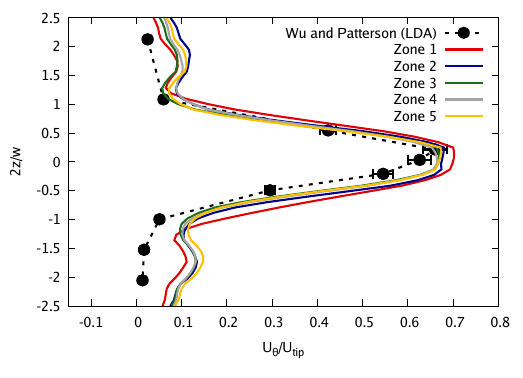}
     \includegraphics[width=0.55\linewidth]{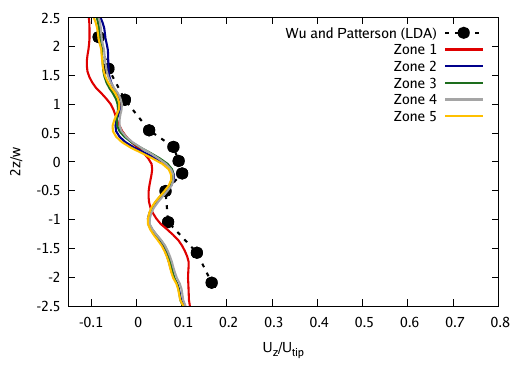}
     \caption{Influence of the MRF zone diameter on the vertical profiles of vertical profiles of velocity components at a distance of 5 cm from the shaft center line: top, radial velocity; middle, tangential velocity; bottom, axial velocity}
     \label{fig:VelocityProfilesMRFZones_5cm}
\end{figure}

\begin{figure}
     \centering
     \includegraphics[width=0.88\linewidth]{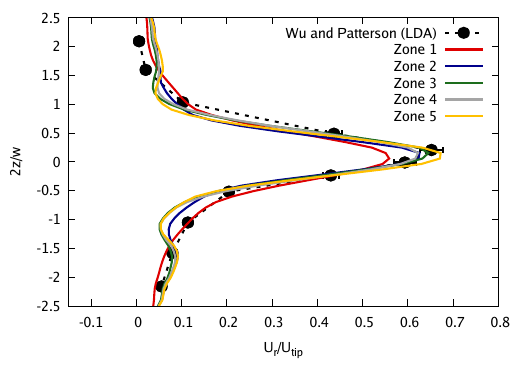}
     \includegraphics[width=0.88\linewidth]{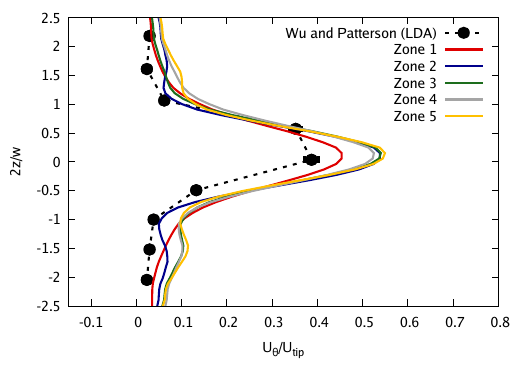}
     \caption{Effect of the MRF zone diameter on the velocity at 6 cm of the shaft axis: top, radial velocity; bottom, tangential velocity}
     \label{fig:VelocityProfilesMRFZones_6cm}
\end{figure}

The impact of the MRF zone diameter on the global performance parameters is shown in Fig \ref{fig:MRFDiameterSensitivityPerfParam}. A sharp increase of more than 10\% in power number and agitation index can be seen between Zone 1 and Zone 2, while a slightly more gradual and linear increase occurs for the turbulence intensity in the tank with a maximum variation of 19\% between Zone 1 and Zone 5. Maximum values are reached with Zone 4 for power number and agitation index.
The radial, tangential and axial velocity profiles at 5 cm from the shaft obtained with the five different MRF zones are shown in Fig. \ref{fig:VelocityProfilesMRFZones_5cm}. The radial velocity prediction is consistent with the experiments of \citet{wu_laser-doppler_1989} and although slight variations in the profile are observed for the Zone 1, the differences in radial velocity when changing the MRF zone diameter are low. Similar observations can be made for the tangential and axial velocity components, except some larger differences occurring between the Zone 1 and the other zones.
As shown in Fig. \ref{fig:VelocityProfilesMRFZones_6cm}, when moving farther away, at 6 cm from the impeller, the smallest MRF region (Zone 1) gives lower peak values for the radial velocity and tangential velocities, thus being the only zone underpredicting the radial velocity peak but, on the other hand, not overshooting the tangential velocity peak as much as the others zones. 

\begin{figure}[ht]
     \centering
     \includegraphics[width=1.0\linewidth]{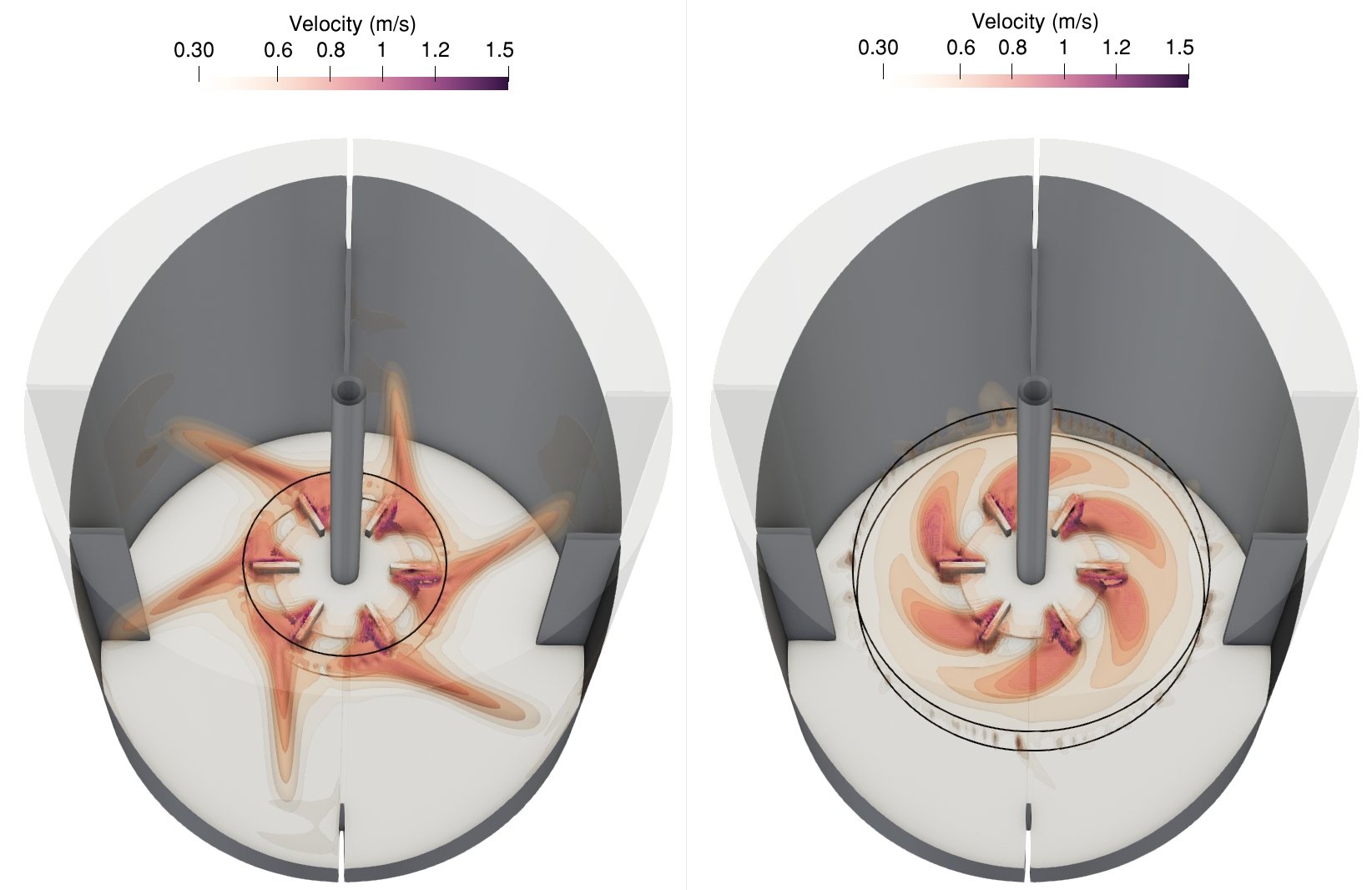}
     \caption{Velocity contours from 0.3 to 1.5 m/s for Zone 1 and Zone 5. The boundary of the MRF zone is shown in black lines: left, Zone 1; right, Zone 5}
     \label{fig:MRFVelocityContours}
\end{figure}

Even though circumferentially-averaged velocity profiles near the impeller blades show limited differences between the different zones, the global velocity field obtained shows some different features, as shown in Fig. \ref{fig:MRFVelocityContours}. The flow field computed with Zone 1 displays a radial distribution of the high velocity stream generated by the blades, which results from the absence of additional rotational velocity term when the flow moves outside of the MRF region. On the other hand, Zone 5 shows the high velocity stream bending when moving away from the impeller. Without a consistent reference data set for the velocity field, making a definitive assertion regarding which of the two zones produces more realistic results is challenging. However, when observing the results of past experimental studies \citep{vant_riet_behaviour_1973, stoots_mean_1995, ng_assessment_1998} and numerical simulations \citep{yeoh_numerical_2004, derksen_large_1999, brucato_numerical_1998, delafosse_trailing_2009, kuschel_validation_2021, ranade_cfd_2002, zadravec_influence_2007}, the flow features of a standard Rushton turbine seem to align more closely with those computed with Zone 5. The velocity field distribution obtained with the smaller zones may be influenced by the fact that they do not strictly satisfy the suggestions of \citet{jaszczur_general_2020} and \citet{zadravec_influence_2007}, as discussed in the introduction. On the other hand, all the MRF zones show good agreement with the reference experimental data, Zone 1 even being closer to the reference data in some of the profiles.

Additionally, one can observe small non-physical velocity variations on the MRF zone boundary for both Zone 1 and 5, although they are greater with Zone 5. The mesh elements orientation with respect to the MRF interface normal direction may impact these spurious velocities, as this can lead to making interpolation errors.

\begin{figure}[ht]
     \centering
     \includegraphics[width=1.0\linewidth]{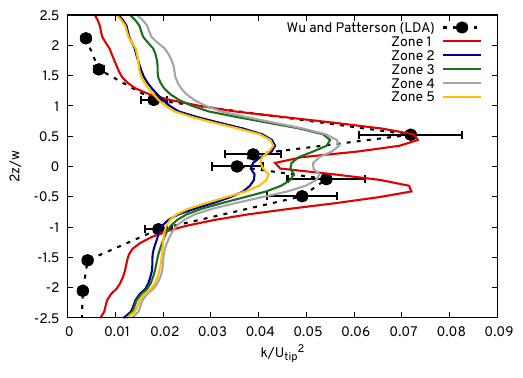}
     \caption{Effect of the MRF zone diameter on vertical profiles of turbulent kinetic energy at 5 cm from the shaft axis}
     \label{fig:TKEProfilesMRFZones}
\end{figure}

\begin{figure}[ht]
     \centering
     \includegraphics[width=1.0\linewidth]{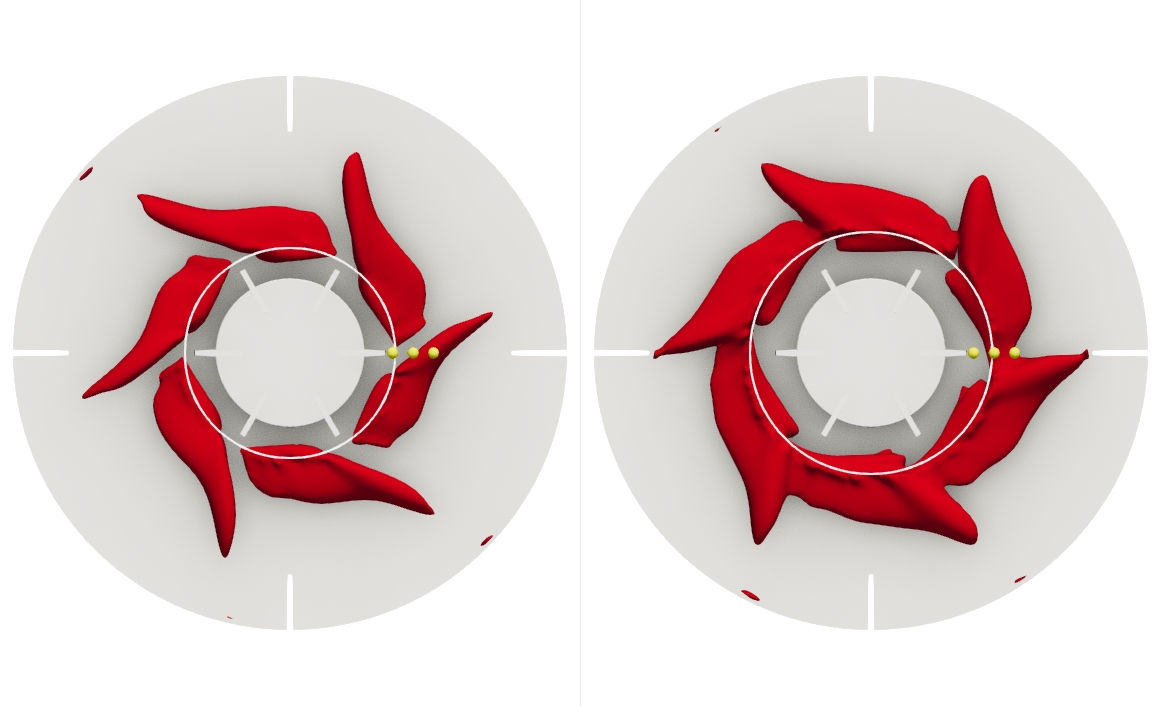}
     \caption{High turbulence intensity zones (above 20\%) for Zone 1 and Zone 2. The boundary of the MRF zone is shown in white, the three probing locations (5, 6 and 7 cm) are shown as black points: left, Zone 1; right, Zone 2}
     \label{fig:TKEZone1vS2}
\end{figure}

The profiles computed for the turbulent kinetic energy using the different MRF zones are presented in Fig. \ref{fig:TKEProfilesMRFZones}, where Zone 1 yields a closer prediction to reference data than the other zones. The decrease of turbulence above and below the main peaks is overpredicted with all the zones, although Zone 1 shows a notable better agreement. Nonetheless, the predicted TKE peak below the impeller generated by the lower part of the blades has the same order of magnitude than the upper peak, while in the data of \citet{wu_laser-doppler_1989} it is approximately 20\% lower. However, since the reference data do not include enough measurement points near the peaks, the lower peak location can potentially be slightly misplaced, thus resulting in a lower maximum value.
Moreover, an unexpected non-monotonous evolution of the profile is observed, with Zone 2 and 5 showing the lowest peak values, while Zone 3 and 4 have substantially higher peak values. In fact, the two most different TKE profiles are those of Zone 1 and Zone 2 which are the zones with the lowest geometrical differences. 
This result can be explained by the proximity of Zone 1 with the 5 cm probing location where $k$ profiles are extracted, while for the other zones this location is at a gradually greater distance from the MRF boundary as shown in Fig. \ref{fig:TKEZone1vS2}. Additionally, depending on where the MRF boundary is located with respect to the main flow features generated by the blades, the high velocity and turbulence zones differs and thus is expected to be the cause of the non-monotonous evolution of the profiles. 

\begin{figure}
     \centering
     \includegraphics[width=0.33\linewidth]{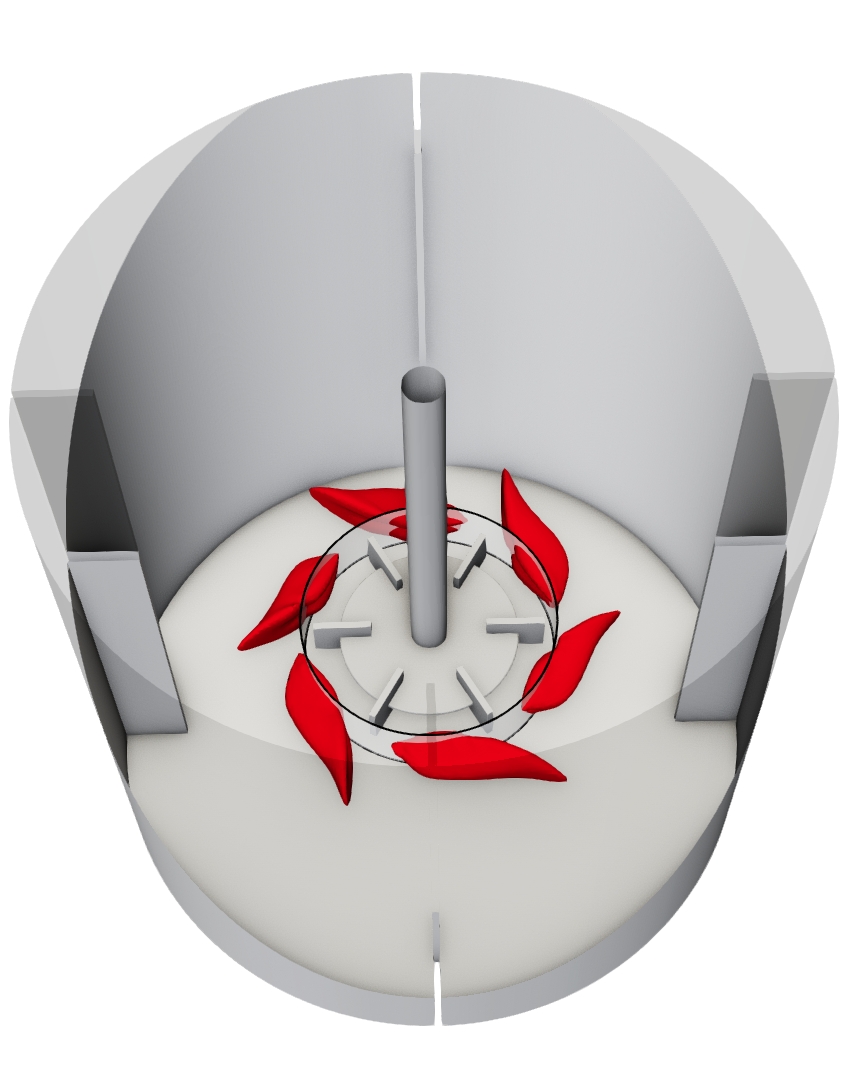}
     \includegraphics[width=0.33\linewidth]{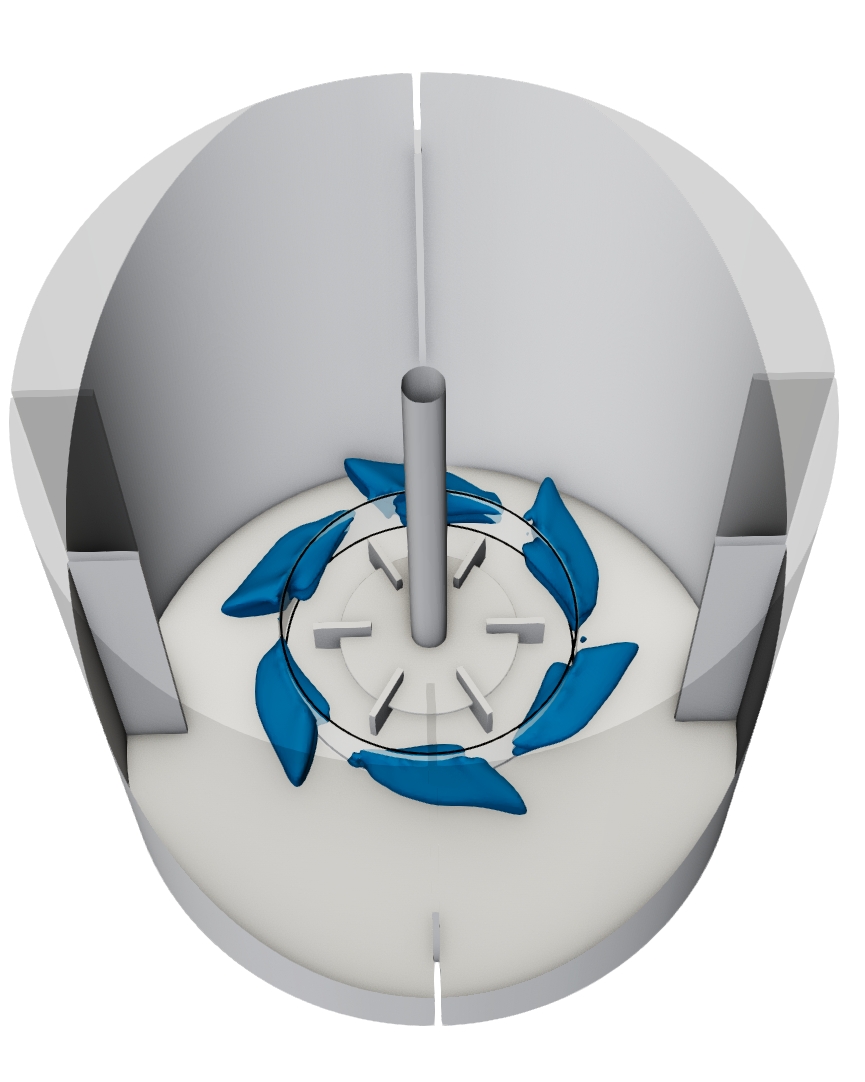}
     \includegraphics[width=0.33\linewidth]{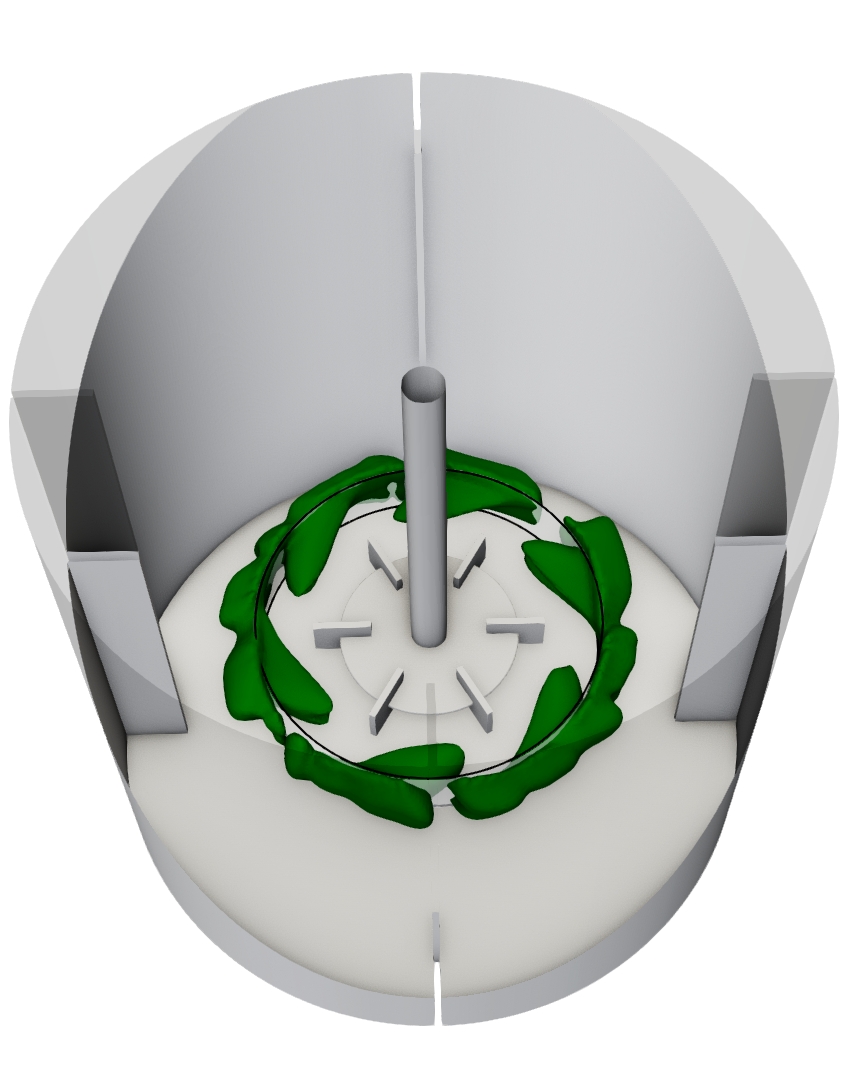}
     \includegraphics[width=0.33\linewidth]{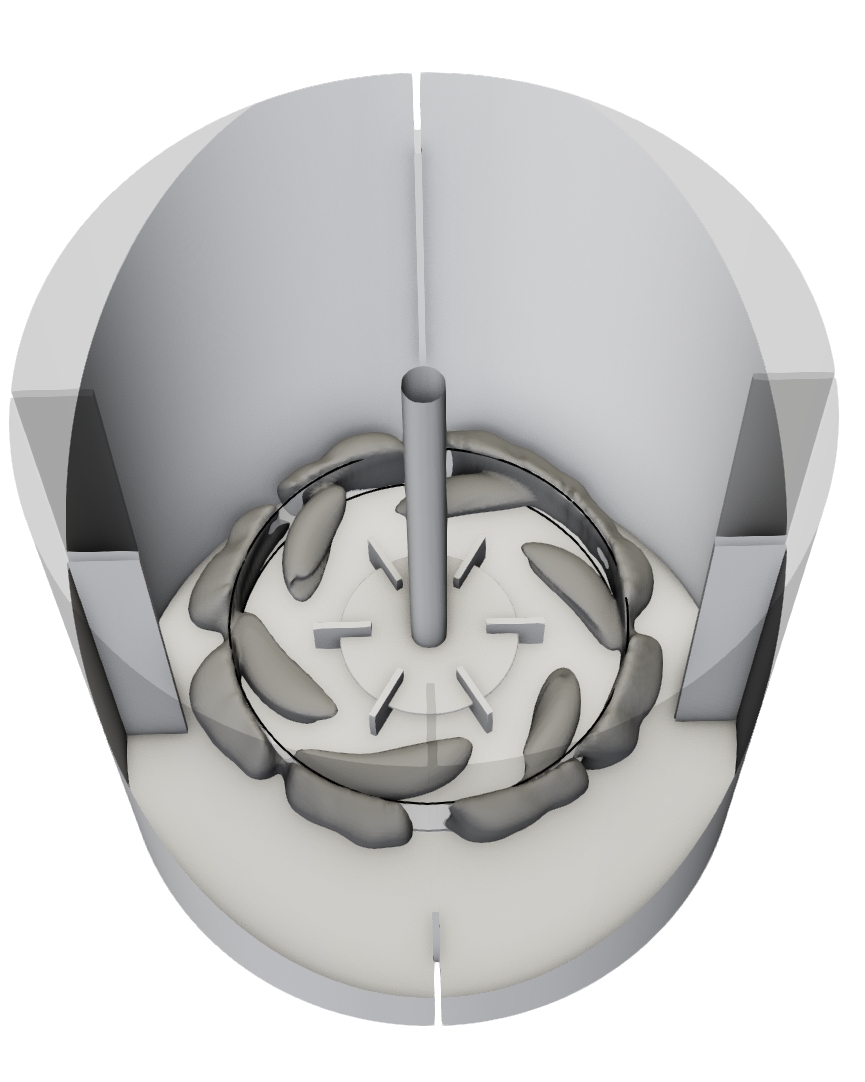}
     \includegraphics[width=0.33\linewidth]{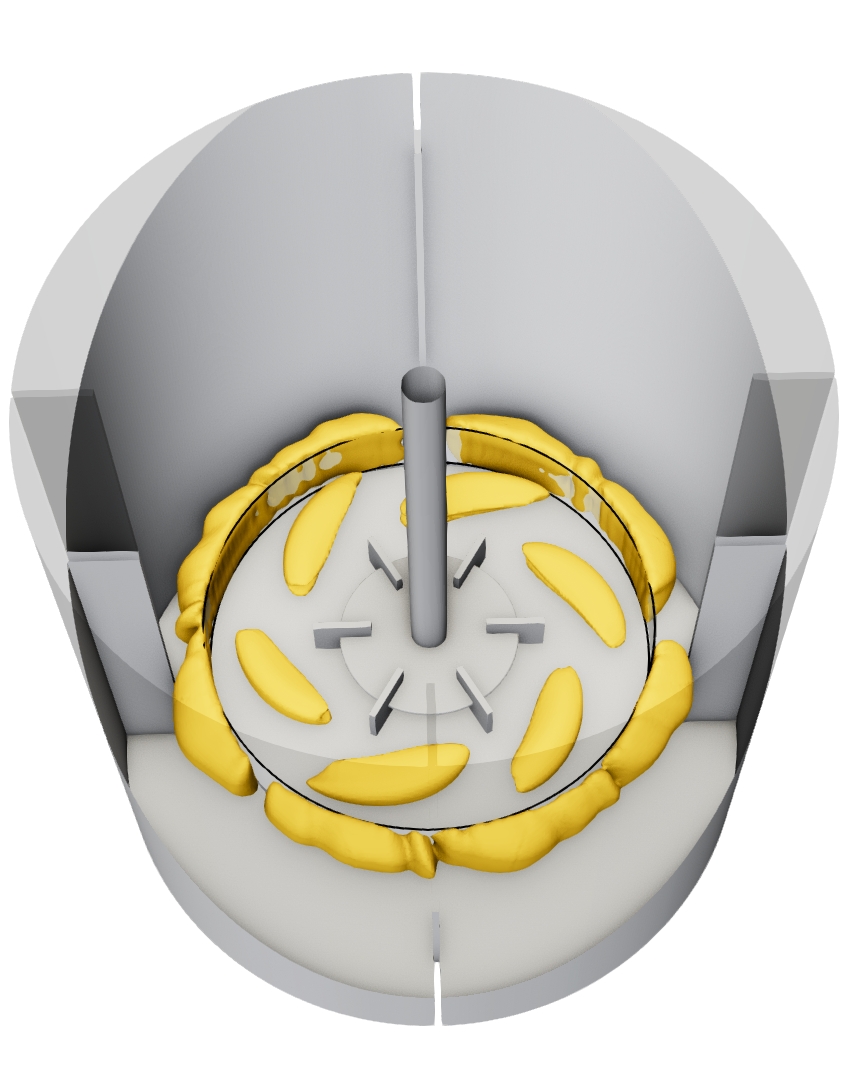}
     \caption{Impact of the MRF zone diameter on the high turbulence intensity zones (over 20\%): red, Zone 1; blue, Zone 2; green, Zone 3; grey, Zone 4; yellow, Zone 5}
     \label{fig:TKEClipMRFZoneComparison}
\end{figure}

\begin{figure}[ht]
     \centering
     \includegraphics[width=1.0\linewidth]{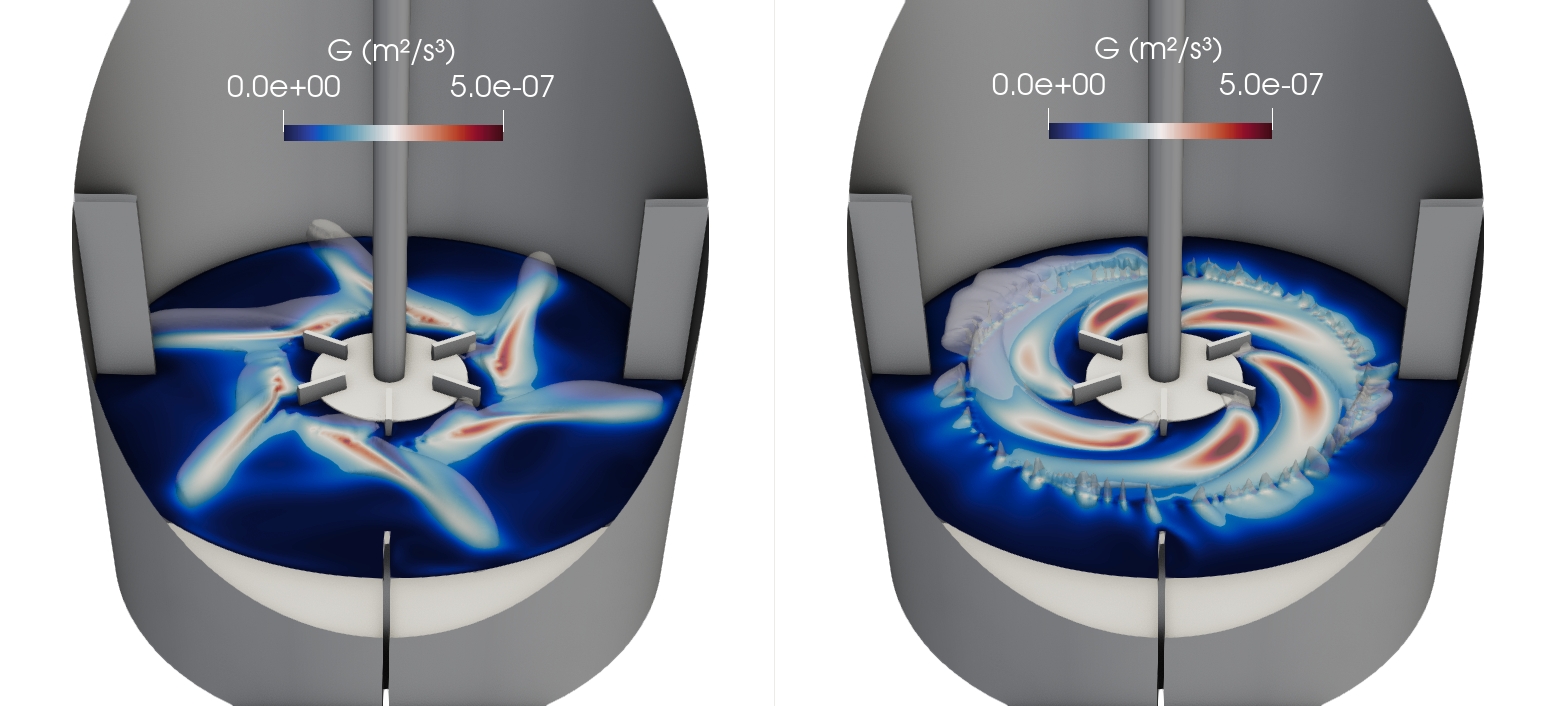}
     \caption{Impact of the MRF zone on the production term $G$ in the impeller plane. Transparent contours represent isosurface of $G=1e-7$ $m^2/s^3$: left, Zone 1; right, Zone 5}
     \label{fig:G_smallVSLarge}
\end{figure}

Another major effect associated with the size of the MRF region is the creation of an artificial turbulent kinetic energy region at the boundary of the MRF zone. Fig. \ref{fig:TKEClipMRFZoneComparison} shows the flow volume with turbulence intensity levels over 20\% for Zone 1, 2, 4 and 5. From Zone 3 to Zone 5, a gradually larger and more defined outskirt of high turbulent kinetic energy is present at the boundary of the MRF zone. To the knowledge of the authors, this behavior has not yet been reported in the literature. In the RANS formulation of Eq. (\ref{eq: kEquationTurbulenceModel}), the turbulent kinetic energy production is based on the Reynolds stress tensor and strain rate tensor components as described in Eq. (\ref{eq: generationTerm}), which depends on the velocity gradients. In the rotating frame, additional momentum terms are added and thus the larger the zone the larger the added term since the distance to the rotation axis is greater. Also, because of the proximity of the MRF zone boundary to the tank wall and baffles spurious velocity jumps can appear at the interface, especially for larger zones as shown in Fig. \ref{fig:MRFVelocityContours}. This effect of the velocity gradient on the turbulent kinetic energy can be observed in Fig. \ref{fig:G_smallVSLarge}, where artificially-created spikes of turbulent kinetic energy production term $G$ are obtained at the interface of Zone 5.
This phenomenon can be highly problematic when the prediction of turbulence levels in the tank is important, and thus for mixing processes where turbulence is a key-factor in promoting mixing. In such cases, a larger MRF zone with an artificial outskirt of high turbulent kinetic energy could numerically reduce the mixing time. Additionally, the turbulent shear stresses play a critical role in certain bioprocesses where the integrity of cell cultures is of paramount importance. For such applications, smaller MRF diameters may be more suitable since it accurately predicts the velocity profiles and turbulent kinetic energy profiles without creating an artificial ring of high turbulent kinetic energy around the MRF boundary. However, this comes with the drawback of potentially impacting the spatial flow distribution for velocities as discussed previously.

\begin{figure}
     \centering
     \includegraphics[width=0.8\linewidth]{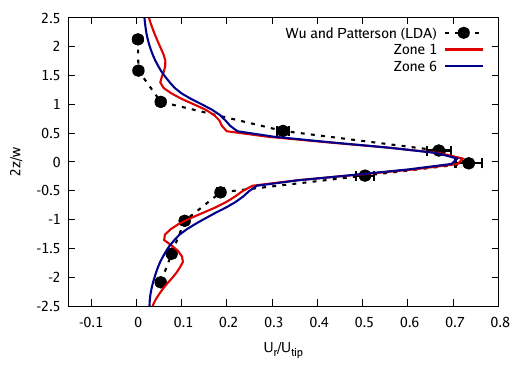}
     \includegraphics[width=0.8\linewidth]{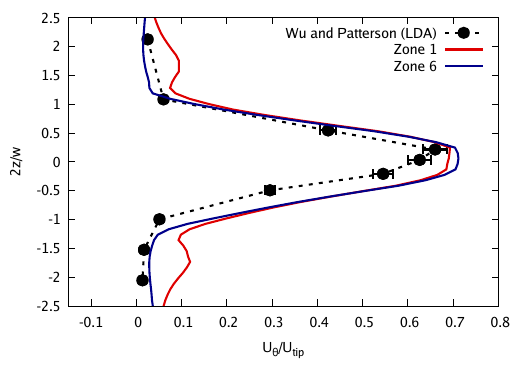}
     \caption{Impact of the MRF zone thickness on the vertical velocity profiles at 5 cm of the shaft axis: top, Zone 1; bottom, Zone 6}
     \label{fig:VelocityComprisonThicknessMRF}
\end{figure}

\begin{figure}
     \centering
     \includegraphics[width=1.0\linewidth]{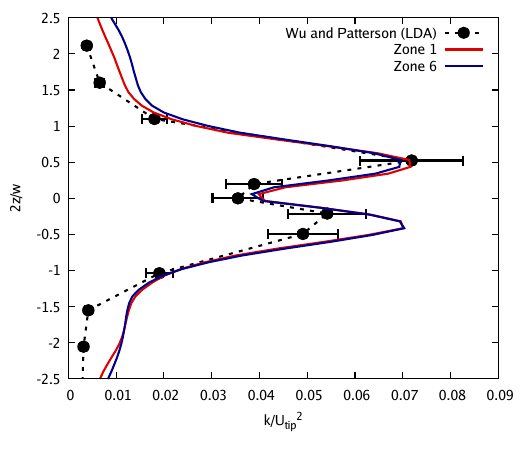}
     \caption{Impact of the MRF zone thickness on the turbulent kinetic energy profiles at 5 cm of the shaft axis}
     \label{fig:TKEComprisonThicknessMRF}
\end{figure}

Following the diameter sensitivity, the thickness of the MRF zone is investigated by comparing Zone 1 to Zone 6, which have the same diameter but respective thicknesses of 1.55\textit{W} and 3.1\textit{W}. For this purpose, the region of the mesh that is refined has been increased vertically in order to encompass the thicker MRF zone. Fig. \ref{fig:VelocityComprisonThicknessMRF} shows the largest differences between the two zones in terms of velocity, with smoother profiles above and below the impeller center line predicted with Zone 6. Overall, there are no significant difference for velocity profiles in the peak region, but there are slight variations in radial and tangential velocities near the blades corners at 5 cm away from the impeller rotating axis.
Similarly, very little differences are observed for the turbulent kinetic energy profiles when changing the MRF zone thickness. The main differences are found near the blades top corner (in the highest turbulence zone) where the thicker zone predicts slightly higher TKE values, as well as one blade width above the impeller, where Zone 1 yields lower TKE values and gives a closer match to the reference data, as shown in Fig. \ref{fig:TKEComprisonThicknessMRF}.

\subsubsection{Impact on mixing}
\label{sec:mixingSensitivity}
In this section, the impact of the size of the MRF region on the mixing in the stirred-tank is investigated by modeling the release of a passive scalar and by computing and comparing the mixing time associated with the calculated flow field using different MRF zones. The following form of the scalar transport equation is solved using a custom solver adapted from the native pimpleFoam solver: 
\begin{equation}
\label{eq:scalarTransportEquation}
    \frac{\partial C}{\partial t} + \nabla \cdot (\mathbf{U}C) = \nabla \cdot \left[ D_{eff} \nabla C \right]
\end{equation}
where $C$ is the scalar concentration and the $D_{eff}$ effective diffusivity, defined as:
\begin{equation}
\label{eq:effectiveDiffusivity}
D_{eff}=D_l+D_t
\end{equation}
where $D_l$ is the laminar diffusivity, $D_t$ is the turbulent diffusivity defined using the turbulent Schmidt number $Sc_t$ as:
\begin{equation}
    D_t=\nu_t/Sc_t
\end{equation}
Typically, it is difficult to pre-determine $Sc_t$ values able to represent the mixing process occurring in different applications and geometries \citep{idzakovicova_mixing_2023}. For example \citet{gualtieri_values_2017} noted that some studies used values ranging from 0.3 to 1. However, given the comparative character of the present analysis, the most common value of $Sc_t=0.7$ adopted in the literature \citep{coroneo_cfd_2011, duan_numerical_2019, hartmann_mixing_2006, gualtieri_values_2017} has been employed. 

The transport equation (\ref{eq:scalarTransportEquation}) is solved by adopting the frozen-flow hypothesis, where the pre-computed velocity field is used to model the convective transport of the passive scalar. This flow/tracer uncoupling method was used by \citet{bujalski_influence_2002} and \citet{duan_numerical_2019}, in the former paper the authors compared this approach with a coupled approach and found little differences in the predicted mixing times. The main discrepancies found in their work were due to the tracer feed position. Nevertheless, this technique remains recognized as a viable method for assessing mixing performance in stirred-tank systems, as outlined in chapter 5-6.4.3 of the handbook of Industrial Mixing \citep{marshall_computational_2003}.

\begin{figure}[ht]
     \centering
     \includegraphics[width=0.7\linewidth]{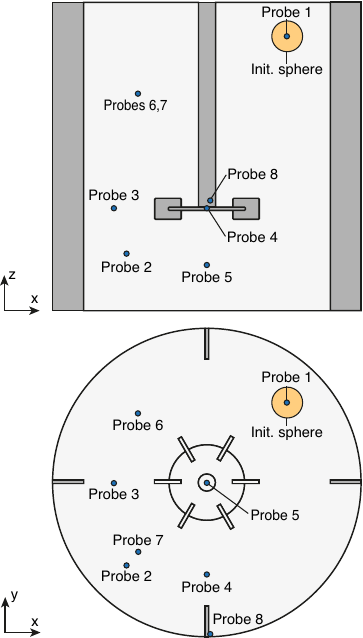}
     \caption{Position of monitoring probes and initialization sphere for the mixing time analysis}
     \label{fig:probesPositionInTank}
\end{figure}

Similarly to \citet{bujalski_influence_2002}, the passive scalar is injected in the tank from the top by initializing the tracer concentration in a sphere near the free surface, as shown in Fig. \ref{fig:probesPositionInTank}. The center of the sphere is located at a radial distance of $\sim$0.1 m from the shaft axis (0.07 m in the x-direction and 0.07 m in the y-direction) and at 0.01 m from the free surface and has a radius of 0.026 m. A series of probes are set to measure the evolution of the concentration of the scalar in time. Their position in the tank is shown in Fig. \ref{fig:probesPositionInTank} and detailed in Table \ref{table:scalarMixingProbingPoints}, where all the distances are specified with respect to the origin point of the reference system. 
\begin{table}[ht!]
\centering
\begin{tabular}{|c|c|c|c|} 
 \hline
 Probe & x (m) & y (m) & z (m)\\ 
 \hline
 1  & 0.07 & 0.07 & 0.15 \\
 2  & -0.07 & -0.07 & -0.04 \\
 3  & -0.08 & 0 & 0 \\
 4  & 0 & 0.08 & 0 \\
 5  & 0 & 0 & -0.05 \\
 6  & -0.06 & 0.06 & 0.1 \\
 7  & -0.06 & -0.06 & 0.1 \\
 8  & 0.003 & -0.133 & 0.007 \\
 \hline
\end{tabular}
\caption{Coordinates of the numerical probes location in the tank}
\label{table:scalarMixingProbingPoints}
\end{table}

The mixing time is defined in a single point location using the 95\% threshold, which is the time after which the normalized scalar concentration $C_{norm}={C(t)}/{C_{\infty}}$ remains within 5\% of its final value, with the final concentration level $C_{\infty}$ defined as: 
\begin{equation}
\label{eq:finalConcentrationValue}
    C_{\infty} = \frac{1}{V_{tank}} \int_{V_{tank}}CdV
\end{equation}
While $C_{norm}$ is used to track the normalized concentration at a single point, the global mixing index $C_{tank}$ can be used to group each probe data into a global concentration for the whole tank. This index is based on the definition of \citet{hartmann_mixing_2006} and slightly modified to better suit a numerical setup. It is used to determine the global mixing time at 95\% of the final concentration and is defined as the variance of the tracer concentrations at all probes location:
\begin{equation}
\label{eq:mixingVarianceTechnique}
    C_{tank}=\frac{1}{n}\sum^{n}_{i=1}\left ({C^i(t) - C_{\infty}} \right )^2
\end{equation}
where $n$ denotes the total number of probes, $C^i(t)$ the concentration at the $i^{th}$ probe location and $C_{\infty}$ the final concentration level in the tank. As the tracer is mixed in the tank, this index decreases and reaches 0 for a perfectly mixed tracer. This method works best with a large amount of probing points, producing in such case an accurate representation of the global mixing quality.

\begin{figure}
     \centering
     \includegraphics[width=0.7\linewidth]{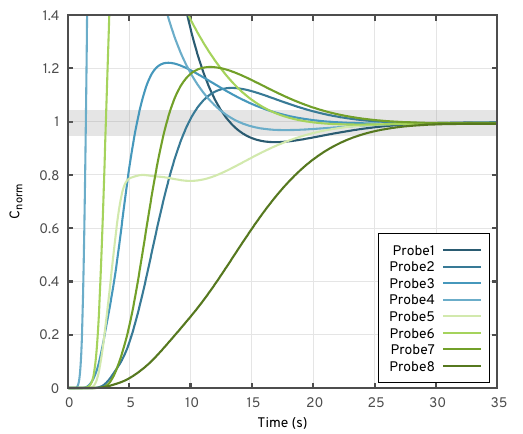}
     \includegraphics[width=0.7\linewidth]{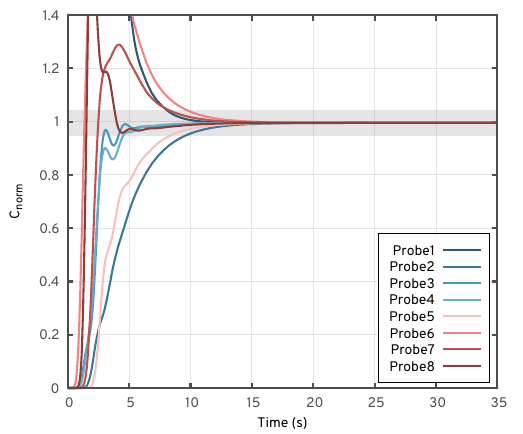}
     \caption{Influence of the MRF zone diameter on the global concentration index $C_{norm}$ of the tracer measured at the monitoring probes: blue to green scale plot, Zone 1; blue to red scale plot, Zone 5; grey area, $\pm$5\% of final value}
     \label{fig:mixingTimeZoneComparison}
\end{figure}

\begin{figure}
     \centering
     \includegraphics[width=1.0\linewidth]{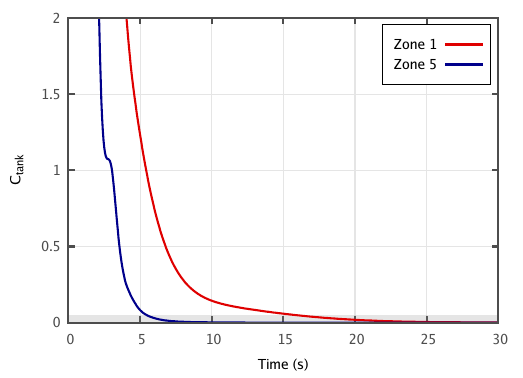}
     \caption{Influence of the MRF zone diameter on the global normalized concentration $C_{tank}$ of the tracer measured at the monitoring probes: blue, Zone 1; red, Zone 5; grey area, $\pm$5\% of final value}
     \label{fig:globalMixingTimeZoneComparison}
\end{figure}

\begin{figure}
     \centering
     \includegraphics[width=1.0\linewidth]{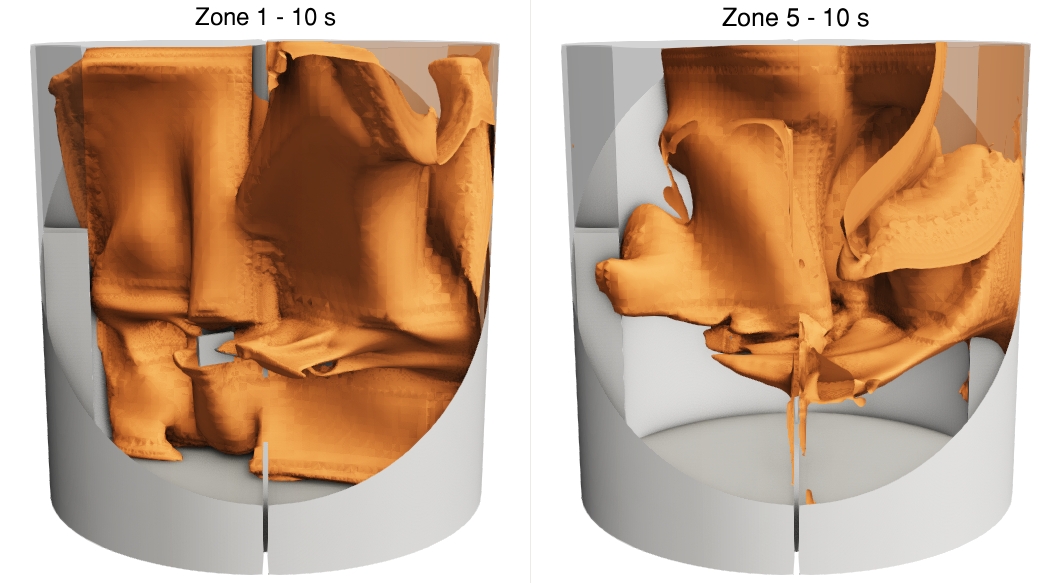}
     \includegraphics[width=1.0\linewidth]{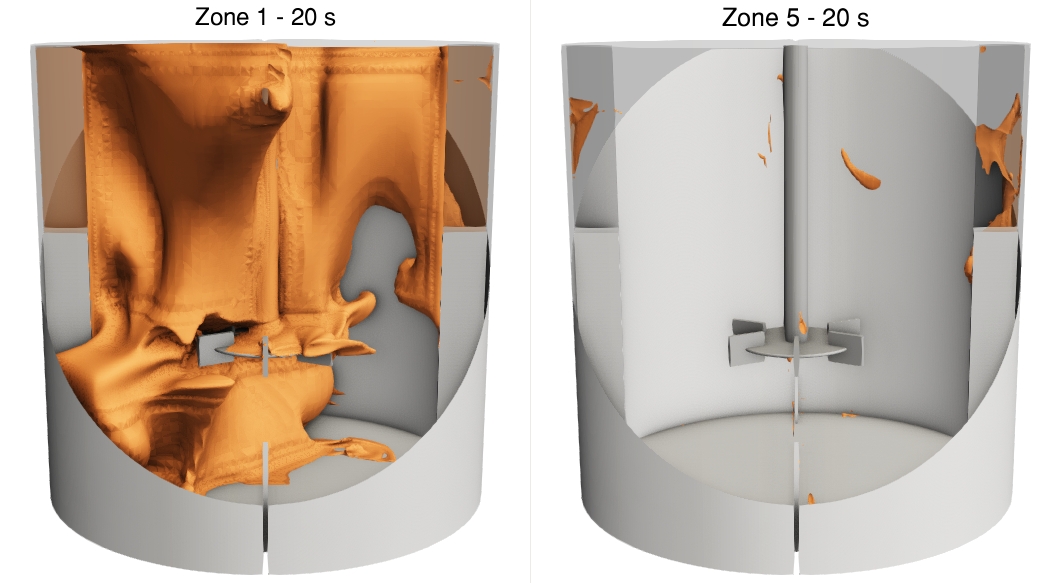}
     \caption{Effect of the MRF zone diameter on the evolution of mixing of the passive tracer. Volume with the tracer concentration over 0.5\% of its equilibrium value at 10 s and 20 s: left, Zone 1; right, Zone 5}
     \label{fig:mixingIsoVolumeComparison}
\end{figure}

When analyzing the MRF zone impact on the tracer mixing, some differences were found in the evolution of single-probe concentrations when changing the zone thickness between Zone 1 and Zone 6; however, the global mixing times obtained were the same. Therefore, the analysis will be focused on the MRF zone diameter only. Fig. \ref{fig:mixingTimeZoneComparison} shows the time evolution of the normalized concentration of the tracer for both MRF Zone 1 (smallest diameter) and Zone 5 (largest diameter), measured at probe 2 (opposite part of the tank compared to the initialization sphere).  
A clear difference is observed between both zones; for instance, the mixing time computed at probe 1 is 9.5 s for Zone 5 and 19.5 s for Zone 1. Thus, in this setup, the use of a larger MRF zone yields local mixing times two to six times lower than when using a smaller MRF zone. This finding could potentially be the reason for the overprediction of mixing time found by \citet{duan_numerical_2019}. Moreover, the tracer concentration evolution at the different probing points is more dampened with the larger zone, and more oscillatory with the smaller zone. The global mixing index $C_{tank}$ is shown in Fig. \ref{fig:globalMixingTimeZoneComparison} for both Zone 1 and Zone 5, resulting in global mixing times of $\theta_{5}=5.5$ s for Zone 5 (largest) and $\theta_{1}=15.9$ s for Zone 1 (smallest). 
The same behavior can be observed in Fig. \ref{fig:mixingIsoVolumeComparison}, where the global mixing is performed under 20 s for Zone 5 (with only residual areas near the baffles having a concentration above 5\% of the equilibrium value). On the other hand, when using Zone 1 there is still a significant part of the tank that is not fully mixed after 20 s.  
Such large differences can be explained by two main factors: increased overall velocities (10\% increase in the agitation index from Zone 1 to Zone 5) and increased turbulence levels when using larger MRF zones (19\% increase in the averaged turbulence intensity from Zone 1 to Zone 5).

\section{Conclusions}
\label{sec:conclusion}
In this study, the MRF technique was used to model the flow in a baffled Rushton turbine with a steady-state approach using the CFD code OpenFOAM. The main objective of the work was to carry out a critical evaluation of the effect of the MRF region size on the predicted flow field and mixing process in the stirred-tank.

Five MRF zone diameters were compared on the basis of global quantities (power number, mixing index and turbulence intensity) and local data (velocity profiles at three radial distances and turbulent kinetic energy profiles near the blades). The smallest zone predicted lower power number than the other zones (more than 12\%), while also predicting lower agitation index and turbulence intensity in the tank. Minor discrepancies were found in the velocity profiles between all the zones, and good agreement with the data of \citet{wu_laser-doppler_1989} was observed overall. 
However, the spatial distribution of high velocity streams differed when changing the MRF zone diameter, possibly due to the MRF interface crossing the blade wake structure. This point remains an open-subject and despite the impact on the spatial velocity distribution, the size of the MRF region had limited effect on the circumferentially-averaged profiles. Spurious high velocity spots were found at the boundary of the MRF region, especially for larger MRF zones. On the other hand, major differences appeared in the turbulent kinetic energy profiles, for which the smallest zone yielded a more accurate prediction of the characteristic double peak profile compared to the other zones. A non-monotonous evolution in the profiles of turbulent kinetic energy was found when increasing the MRF zone diameter, with Zone 1 and Zone 2 showing the largest difference ($\sim$ 85\%) between their TKE peak value with respect to the other zones. Moreover, artificial creation of turbulent kinetic energy was found at the MRF interface, with increasing presence for larger MRF diameters.
To the knowledge of the authors this phenomenon has not yet been reported in the literature and seems to originate from the spurious velocity gradients at the MRF boundary. This artifact becomes larger as the MRF interface gets closer to the baffles and tank walls.

Additionally, Zone 1 was used as a basis to investigate the impact of two different MRF zone thicknesses on the flow. It was shown that this parameter had a limited influence on the global results. However, small differences in velocity and turbulent kinetic energy were found near the upper and lower edges of the blades, where the thicker zone provided slightly closer velocity predictions when compared with the literature data. Nonetheless, the thinner zone yielded better turbulent kinetic energy predictions.

Lastly, unsteady scalar mixing simulations were performed based on the frozen-flow field computed with the smallest and largest MRF zones in order to assess the impact of the MRF region diameter on the mixing time predictions. Large differences in mixing time were found between the two zones, with values ranging from 3.8 s up to 23.6 s when switching from the largest to the smallest region, because of the greater average velocities and artificially generated turbulence associated with the larger one.

The k-$\omega$ SST and k-$\epsilon$ turbulence model were also compared in terms of stability and accuracy, the former was found to be less stable and captured some of the unsteady-based oscillations in the residuals and physical quantities. However, while no major difference was found in the velocity profiles between the turbulence models, the k-$\omega$ SST predicted more accurately the turbulent kinetic energy profiles compared to the k-$\epsilon$, with a maximum difference of almost 80\% in the blade center region.

To conclude, this paper highlighted the impact of the MRF zone dimensions on the stirred-tank simulation accuracy and stresses the need for additional numerical studies based on the MRF method to address the effect of selected the MRF region on numerical results. Similarly to the standard grid sensitivity analysis, the authors believe systematic MRF sensitivity studies will lead to more reliable and standardized practices.

\backmatter








\section*{Declarations}
\subsection*{Acknowledgments}
The authors would like to thank the organizing committee of the $18^{th}$ OpenFOAM Workshop for the valuable discussions on the topic during the conference, which greatly contributed to the preparation of this manuscript

\subsection*{Competing interest}
\begin{itemize}
\item A.B. and C.C. are Chiesi Farmaceutici employees.
\item A.R. and R.R. are consultants/contractors currently engaged by Chiesi Farmaceutici.
\end{itemize}

\subsection*{Author contribution}
A. Reid: data curation, formal analysis, investigation, writing – original draft, writing – review \& editing, visualization.\\ 
R. Rossi: conceptualization, writing – original draft, writing – review \& editing, supervision.\\  
C. Cottini: conceptualization, project administration, resources, supervision, writing – review \& editing. \\ 
A. Benassi: conceptualization, investigation, project administration, resources, supervision, visualization, writing – review \& editing.

\subsection*{Data, material and code availability}
Data supporting this study can be made available upon request. The access is subject to approval and a data sharing agreement.









\begin{appendices}

\section{Blade angular sensitivity}
\label{appendix: bladeAngularSensitivity}
\begin{figure}
     \centering
     \includegraphics[width=0.55\linewidth]{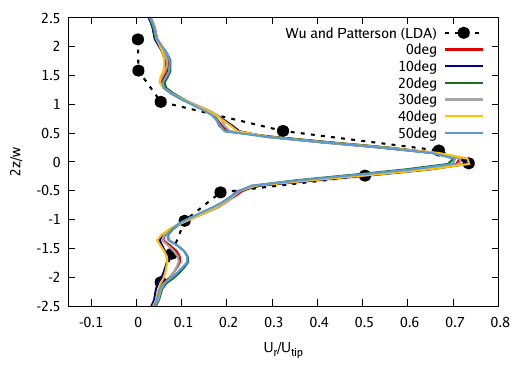}
     \includegraphics[width=0.55\linewidth]{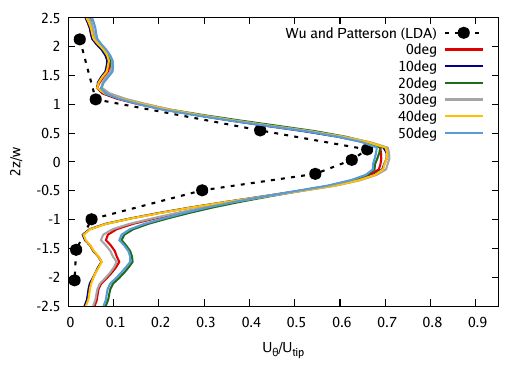}
     \includegraphics[width=0.55\linewidth]{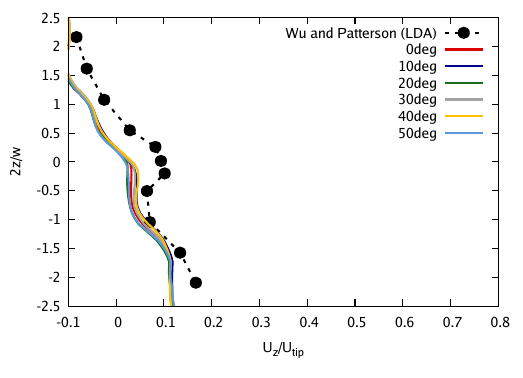}
     \caption{Vertical profiles of velocity components at a distance of 5 cm from the shaft axis for the five tested angular positions of the blades: top, radial velocity; middle, tangential velocity; bottom, axial velocity}
     \label{fig:profilesBladeAngularPosition_5cm}
\end{figure}

\begin{figure}[ht]
     \centering
     \includegraphics[width=0.9\linewidth]{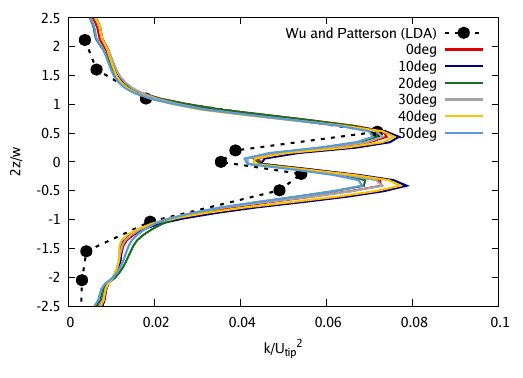}
     \caption{Comparison of vertical profiles of turbulent kinetic energy at a distance of 5 cm from the shaft axis for the five tested angular positions of the blades}
     \label{fig:bladeAngularPositionComparison}
\end{figure}

Simulations have been ran with different impeller positions with respect to the baffles in order to assess the impact on the performance parameters of this study. Five angular position of the blades have been tested, starting from the base position (defined as \textit{0deg}) displayed in Fig. \ref{fig:geometryRushton} and then rotating the impeller by 10° up to 50°. 
Fig. \ref{fig:profilesBladeAngularPosition_5cm} shows the velocity profiles for five different angular positions and similarly to the findings of \citet{harvey_steady-state_1995} only minor differences are observed. On the other hand, Fig. \ref{fig:bladeAngularPositionComparison} shows $k$ profiles at 5 cm radial distance from the shaft center line for the five angular position tested and non-negligible differences can be seen in $k$ peak regions. 




\end{appendices}


\bibliography{references}

\end{document}